\documentclass[12pt,a4paper]{article}
\usepackage{graphics,graphicx}
\usepackage{amssymb,epsfig,amsmath,euscript,array}
\usepackage{cite}
\makeatletter
\@addtoreset{equation}{section}
\makeatother
\renewcommand{\theequation}{\thesection.\arabic{equation}}

\newcommand{\startappendix}{
\setcounter{section}{0}
\renewcommand{\thesection}{\Alph{section}}
\renewcommand{\theequation}{\Alph{section}.\arabic{equation}}}

\newcommand{\Appendix}[1]{
\refstepcounter{section}
\begin{flushleft}
{\Large\bf Appendix \thesection: #1}
\end{flushleft}}

\newcounter{multieqs}

\newcommand{\be}{\begin{equation}}
\newcommand{\ee}{\end{equation}}

\newcommand{\bm}[1]{\mbox{\boldmath $#1$}}

\newcommand{\kslash}{k \!\!\! / }
\newcommand{\qslash}{q \!\!\! / }
\newcommand{\lslash}{l \!\! / }
\newcommand{\Pslash}{P \!\!\!\! / }

\def\bd{\begin{document}}
\def\ed{\end{document}}
\def\nn{\nonumber}
\def\bea{\begin{eqnarray}}
\def\eea{\end{eqnarray}}
\let\bm=\bibitem
\let\la=\label

\def\npb#1#2#3{Nucl. Phys. {\bf{B#1}} #3 (#2)}
\def\plb#1#2#3{Phys. Lett. {\bf{#1B}} #3 (#2)}
\def\prl#1#2#3{Phys. Rev. Lett. {\bf{#1}} #3 (#2)}
\def\prd#1#2#3{Phys. Rev. {D \bf{#1}} #3 (#2)}
\def\cmp#1#2#3{Comm. Math. Phys. {\bf{#1}} #3 (#2)}
\def\cqg#1#2#3{Class. Quantum Grav. {\bf{#1}} #3 (#2)}
\def\nppsa#1#2#3{Nucl. Phys. B (Proc. Suppl.) {\bf{#1A}}#3 (#2)}
\def\ap#1#2#3{Ann. of Phys. {\bf{#1}} #3 (#2)}
\def\ijmp#1#2#3{Int. J. Mod. Phys. {\bf{A#1}} #3 (#2)}
\def\rmp#1#2#3{Rev. Mod. Phys. {\bf{#1}} #3 (#2)}
\def\mpla#1#2#3{Mod. Phys. Lett. {\bf A#1} #3 (#2)}
\def\jhep#1#2#3{J. High Energy Phys. {\bf #1} #3 (#2)}
\def\atmp#1#2#3{Adv. Theor. Math. Phys. {\bf #1} #3 (#2)}
\newcommand{\EQ}[1]{\begin{equation} #1 \end{equation}}
\newcommand{\AL}[1]{\begin{subequations}\begin{align} #1 \end{align}\end{subequations}}
\newcommand{\SP}[1]{\begin{equation}\begin{split} #1 \end{split}\end{equation}}
\newcommand{\ALAT}[2]{\begin{subequations}\begin{alignat}{#1} #2 \end{alignat}
                        \end{subequations}}
\def\beqa{\begin{eqnarray}}
\def\eeqa{\end{eqnarray}}
\def\beq{\begin{equation}}
\def\eeq{\end{equation}}
\def\N{{\cal N}}
\def\sst{\scriptscriptstyle}
\def\thetabar{\bar\theta}
\def\Tr{{\rm Tr}}
\def\one{\mbox{1 \kern-.59em {\rm l}}}
 \def\Nh{\hat{N}}

\def\a{\alpha}      \def\da{{\dot\alpha}}
\def\b{\beta}       \def\db{{\dot\beta}}
\def\c{\gamma}  \def\G{\Gamma}  \def\cdt{\dot\gamma}
\def\d{\delta}  \def\D{\Delta}  \def\ddt{\dot\delta}
\def\e{\epsilon}        \def\vare{\varepsilon}
\def\f{\phi}    \def\F{\Phi}    \def\vvf{\f}
\def\h{\eta}
\def\k{\kappa}
\def\l{\lambda} \def\L{\Lambda}
\def\m{\mu} \def\n{\nu}
\def\o{\omega}
\def\p{\pi} \def\P{\Pi}
\def\r{\rho}
\def\s{\sigma}  \def\S{\Sigma}
\def\t{\tau}
\def\th{\theta} \def\Th{\Theta} \def\vth{\vartheta}
\def\X{\Xeta}
\def\z{\zeta}
\def\cA{{\cal A}} \def\cB{{\cal B}} \def\cC{{\cal C}}
\def\cD{{\cal D}} \def\cE{{\cal E}} \def\cF{{\cal F}}
\def\cG{{\cal G}} \def\cH{{\cal H}} \def\cI{{\cal I}}
\def\cJ{{\cal J}} \def\cK{{\cal K}} \def\cL{{\cal L}}
\def\cM{{\cal M}} \def\cN{{\cal N}} \def\cO{{\cal O}}
\def\cP{{\cal P}} \def\cQ{{\cal Q}} \def\cR{{\cal R}}
\def\cS{{\cal S}} \def\cT{{\cal T}} \def\cU{{\cal U}}
\def\cV{{\cal V}} \def\cW{{\cal W}} \def\cX{{\cal X}}
\def\cY{{\cal Y}} \def\cZ{{\cal Z}}
\def\ua{\underline{\alpha}}
\def\ub{\underline{\phantom{\alpha}}\!\!\!\beta}
\def\uc{\underline{\phantom{\alpha}}\!\!\!\gamma}
\def\um{\underline{\mu}}
\def\ud{\underline\delta}
\def\ue{\underline\epsilon}
\def\una{\underline a}\def\unA{\underline A}
\def\unb{\underline b}\def\unB{\underline B}
\def\unc{\underline c}\def\unC{\underline C}
\def\und{\underline d}\def\unD{\underline D}
\def\une{\underline e}\def\unE{\underline E}
\def\unf{\underline{\phantom{e}}\!\!\!\! f}\def\unF{\underline F}
\def\unm{\underline m}\def\unM{\underline M}
\def\unn{\underline n}\def\unN{\underline N}
\def\unp{\underline{\phantom{a}}\!\!\! p}\def\unP{\underline P}
\def\unq{\underline{\phantom{a}}\!\!\! q}
\def\unQ{\underline{\phantom{A}}\!\!\!\! Q}
\def\unH{\underline{H}}
\def\As {{A \hspace{-6.4pt} \slash}\;}
\def\bs {{b \hspace{-6.4pt} \slash}\;}
\def\Ds {{D \hspace{-6.4pt} \slash}\;}
\def\ds {{\del \hspace{-6.4pt} \slash}\;}
\def\ss {{\s \hspace{-6.4pt} \slash}\;}
\def\ks {{ k \hspace{-6.4pt} \slash}\;}
\def\ps {{p \hspace{-6.4pt} \slash}\;}
\def\pas {{{p_1} \hspace{-6.4pt} \slash}\;}
\def\pbs {{{p_2} \hspace{-6.4pt} \slash}\;}
\def\Fh{\hat{F}}
\def\Vh{\hat{V}}
\def\Xh{\hat{X}}
\def\ah{\hat{a}}
\def\xh{\hat{x}}
\def\yh{\hat{y}}
\def\ph{\hat{p}}
\def\xih{\hat{\xi}}
\def\psit{\tilde{\psi}}
\def\Psit{\tilde{\Psi}}
\def\tht{\tilde{\th}}
\def\lt{\tilde{\lambda}}
\def\llt{\tilde{l}}
\def\At{\tilde{A}}
\def\Qt{\tilde{Q}}
\def\Rt{\tilde{R}}
\def\Nt{\tilde{N}}

\def\at{\tilde{a}}
\def\st{\tilde{s}}
\def\ft{\tilde{f}}
\def\pt{\tilde{p}}
\def\qt{\tilde{q}}
\def\vt{\tilde{v}}
\def\nt{\tilde{n}}
\def\delb{\bar{\partial}}
\def\bz{\bar{z}}
\def\bD{\bar{D}}
\def\bB{\bar{B}}
\def\bk{{\bf k}}
\def\bl{{\bf l}}
\def\bp{{\bf p}}
\def\bq{{\bf q}}
\def\br{{\bf r}}
\def\bx{{\bf x}}
\def\by{{\bf y}}
\def\bR{{\bf R}}
\def\bV{{\bf V}}
\def\d{\delta}\def\D{\Delta}\def\ddt{\dot\delta}
\def\pa{\partial} \def\del{\partial}
\def\xx{\times}
\def\uno{\mbox{1 \kern-.59em {\rm l}}}
\def\trp{^{\top}}
\def\inv{^{-1}}
\def\dag{{^{\dagger}}}
\def\pr{^{\prime}}
\def\lan{\langle}
\def\ran{\rangle}
\def\rar{\rightarrow}
\def\lar{\leftarrow}
\def\lrar{\leftrightarrow}
\newcommand{\0}{\,\!}      %this is just NOTHING!
\def\one{1\!\!1\,\,}
\def\im{\imath}
\def\jm{\jmath}
\newcommand{\tr}{\mbox{tr}}
\newcommand{\slsh}[1]{/ \!\!\!\! #1}
\def\vac{|0\rangle}
\def\lvac{\langle 0|}
\def\hlf{\frac{1}{2}}
\def\ove#1{\frac{1}{#1}}
\def\Box{\square}
\def\ZZ{\mathbb{Z}}
\def\CC#1{({\bf #1})}
\def\bcomment#1{}
\def\bfhat#1{{\bf \hat{#1}}}
\def\VEV#1{\left\langle #1\right\rangle}
\newcommand{\ex}[1]{{\rm e}^{#1}} \def\ii{{\rm i}}
\def\rr{{\rm r}} \def\rs{{\rm s}}\def\rv{{\rm v}}
\def\ri{{\rm i}}\def\rj{{\rm j}}
\newcommand{\lrbrk}[1]{\left(#1\right)}
\newcommand{\sfrac}[2]{{\textstyle\frac{#1}{#2}}}

\def\Li2{{\rm Li}_2}

\font\mybb=msbm10 at 12pt
\def\bb#1{\hbox{\mybb#1}}

\font\myBB=msbm10 at 18pt
\def\BB#1{\hbox{\myBB#1}}

\begin{document}

\begin{flushright}
hep-th/0410280 \\
QMUL-PH-04-09
\end{flushright}

\vspace{20pt}

\begin{center}

{\Large \bf A Twistor Approach to One-Loop Amplitudes  } \\
\vspace{12pt}
{\Large \bf  in $\cN   \! = \!  1$ Supersymmetric  Yang-Mills Theory
  \\}
\vspace{33pt}

{\bf James Bedford, Andreas  Brandhuber, Bill  Spence and Gabriele  Travaglini}
\footnote{
{\sffamily \{\tt j.a.p.bedford, a.brandhuber, w.j.spence,
g.travaglini\}@qmul.ac.uk }}

{\em Department of Physics\\
Queen Mary, University of
London\\
Mile End Road, London, E1 4NS\\
United Kingdom
 }

\vspace{40pt} {\bf Abstract}

\end{center}

% ABSTRACT goes here

\noindent
We extend the twistor string theory inspired 
formalism introduced in hep-th/0407214 for 
calculating loop amplitudes in $\cN  \! = \! 4$ 
super Yang-Mills theory to the case of  $\cN \! = \! 1$ 
(and $\cN \! = \! 2$)  super Yang-Mills. 
Our approach yields a novel representation of the gauge theory 
amplitudes as dispersion integrals, which are surprisingly simple
to evaluate. 
As an application we calculate one-loop maximally helicity 
violating (MHV)  scattering amplitudes with an arbitrary 
number of external legs. 
The result we obtain agrees precisely with the 
expressions for the $\cN \! = \! 1$ MHV amplitudes derived  
previously by Bern, Dixon, Dunbar and Kosower 
using the cut-constructibility approach.

\vspace{0.5cm}

\setcounter{page}{0}
\thispagestyle{empty}
\newpage

%%%%%%%%%%%%%%%%%%%%%%%%%%%%%%%%%%%%%%%%%%%%%%%%%%%%%%%%%%%%%%%%%
%%%%%%%%%%%%%%%%%%%%%%%%%%%%%%%%%%%%%%%%%%%%%%%%%%%%%%%%%%%%%%%%%
%%%%%%%%%% ordinary document (end)%%%%%%%%%%%%%%%%%%%%%%%%%%%%%%%

\section{Introduction}

In a remarkable paper \cite{witten} it was proposed that perturbative
$\cN=4$ super Yang-Mills (SYM) is dual to the topological B model on
super twistor space $\bb{C}\bb{P}^{\rm 3|4}$. Interestingly, this duality
relates the perturbative expansion of gauge theory amplitudes to a
D1-brane instanton expansion on the string theory side. The relevant 
instantons
correspond to algebraic curves embedded holomorphically in super twistor
space. Their degree $d$ and genus $g$ are related to the number of
negative helicity gluons $q$ and number of loops $l$ of the amplitude as
$d=q-1+l$ and $g \leq l$.

In \cite{witten} the maximally helicity violating
(MHV) tree amplitudes \cite{mhv,bg} were reproduced directly from a
computation in the B model with $d=1$, $g=0$ curves. 
For more general amplitudes the story becomes more involved,
since in 
\cite{witten} it was already pointed out that in principle both
connected and disconnected instantons can contribute to 
a given amplitude. The approach using connected instantons
was pursued further in \cite{rsv,rv,rsv2}, and agreement with
existing results was found. On the
other hand,  \cite{csw} introduced a new diagrammatic method to
calculate tree-level amplitudes using MHV amplitudes as 
effective vertices after a suitable off-shell continuation. 
This method is related to the string theory approach 
using contributions from completely disconnected instantons only, 
and is extremely efficient for calculating tree amplitudes. 
The authors of \cite{gmn} argued that both computations are equivalent,
and that the instanton contributions localize on singular curves
corresponding to intersecting ``lines'', i.e.~degree one curves 
in twistor space. 

The  method  of \cite{csw} led to a rederivation of previously known 
results in a much faster way, and also allowed the calculation of 
new scattering amplitudes with increasing helicity violation 
as well as with fermionic and scalar external particles 
\cite{Zhu}--\!\!\cite{vvkreview}. 
The next logical question to ask was clearly 
whether this procedure could be extended to 
the calculation of amplitudes at loop level. 
This question was positively answered by three of the present authors 
in \cite{bst}, where it was shown how to combine 
MHV vertices into loop diagrams in $\cN=4$ SYM theory.
The result was a new representation of the 
scattering amplitudes in terms of dispersive integrals 
which, rather surprisingly, proved to be tractable, and 
% The explicit calculations carried out in \cite{bst}   
led to perfect agreement with the expressions 
for the one-loop MHV scattering amplitudes in $\cN=4$ SYM 
previously obtained by Bern, Dixon, Dunbar and Kosower 
(BDDK) in \cite{Bern:zx} using 
the cut-constructibility approach. 
As a bonus, the analysis of \cite{bst} gave a 
novel representation of the
``easy two-mass''  (2me) scalar box function, 
containing one less dilogarithm and 
one less logarithm than the traditional one of 
BDDK.

The twistor string theory originally proposed in \cite{witten} 
and other versions proposed in \cite{Berkovits, BerkMotl}
cannot reproduce gauge theory amplitudes 
at loop level \cite{BerkWitt2}, due to the fact 
that conformal supergravity fields propagate in the loops.
Nevertheless, one can study known loop amplitudes to
extract information about their localisation properties in twistor space.
This was done in \cite{csw2} for supersymmetric MHV one-loop
amplitudes (and some non-supersymmetric amplitudes), and
localisation was indeed found onto three types of diagrams.
The twistor space picture suggested by the result 
of \cite{bst} differed from that of \cite{csw2}, 
in that one of the class of diagrams
proposed in the analysis of [22] is absent. 
The resolution of this discrepancy was recently \cite{csw3} linked
to a subtlety in the use of the differential operators 
employed to establish localisation, and led to the appearance of a
\lq\lq holomorphic anomaly\rq\rq . 
With this taken into account, 
it was shown in \cite{csw3} that, 
as far as the one-loop $\cN=4$ SYM amplitudes
are concerned, the expectations  
from \cite{bst} are confirmed and the
one-loop MHV amplitudes in $\cN=4$
localise on pairs of lines in twistor space which are 
joined by two twistor space propagators. 
The holomorphic anomaly was evaluated explicitly in \cite{benabern, freddy},  
and twistor space localisation was further elucidated in \cite{benabern} by 
transforming the results of \cite{csw} and \cite{bst} directly to twistor space.
Furthermore, the holomorphic anomaly was recently exploited  
as a new tool for deriving one-loop next-to-MHV amplitudes in $\cN=4$ SYM
\cite{freddy,freddy2} in combination with the powerful 
cut-constructibility approach. These and other new one-loop amplitudes in
$\cN=4$ SYM were recently derived in \cite{new} using  cut-constructibility.

It now appears plausible that the entire quantum theory 
of $\cN=4$ SYM will have a description based upon the 
MHV diagram approach, which in turn reflects the properties 
of localisation in twistor space.
Moreover, we now have direct methods to
test this, using MHV vertices assembled into
MHV diagrams according to well-defined rules.
We expect further evidence to emerge with the continuing
study of non-MHV tree and one-loop amplitudes,
as well as higher-loop amplitudes.
It should be fairly straightforward to check
that the few remaining existing
results that have so far not
been obtained by these new methods are reproduced.
%The improved efficacy of these
%methods has moreover already enabled to derive additional
%new results \cite{freddy,freddy2,new}.

We find it intriguing that, although 
the twistor string/gauge theory correspondence 
is spoiled at loop level, one can still use MHV vertices 
and combine them to obtain the correct $\cN=4$ SYM loop amplitudes \cite{bst}. 
Perhaps even more surprisingly, we will show in this paper that 
the applicability of the method of \cite{bst} 
actually holds for theories with less supersymmetry as well, 
which   are interesting  not least for 
potential applications to phenomenology.  
Specifically, in this paper we will obtain  
one-loop MHV amplitudes in  $\cN=1$ SYM  
by combining MHV vertices into one-loop diagrams following 
the procedure proposed in \cite{bst}.   
The result we find agrees perfectly 
with the BDDK computation of \cite{Bern:1994cg}.%
\footnote
{The arguments presented here
also apply directly to 
$\cN=2$ SYM.}
% and this
%will be understood in the following.}

The rest of this paper is organised as follows. 
In Section 2 we begin by reviewing the expression 
for the one-loop MHV scattering amplitudes
in $\cN=1$ SYM derived by BDDK in 
\cite{Bern:1994cg}. There, we also present 
a slightly simplified version of the BDDK result 
which will be useful in making the comparison 
with the results derived from MHV vertices.
Section 3 reviews relevant aspects of the work of
\cite{bst},  thereby establishing the set-up for 
the calculation of loop amplitudes with MHV vertices.  
In Section 4 we turn to the formulation
and explicit calculation of the one-loop 
MHV amplitudes in $\cN=1$ SYM.
Finally, in three Appendices we summarise 
some technical results which are required 
in the course of the calculations presented in 
Section 4. 
%This work requires some technical results
%which are summarised in three Appendices.

%%%%%%%%%%%%%%%%%%%%%%%%%%%%%%%%%%%%%%%%%%%%%%%%%

\section{The $\cN=1$ MHV amplitudes at one loop}
The expression for the MHV amplitudes
at one loop in $\cN=1$ SYM was obtained for
the first time by BDDK in \cite{Bern:1994cg},
using the cut-constructibility method.
We will shortly give their explicit result,
and then simplify it
by introducing appropriate functions. This
turns out to be useful when we compare
the BDDK result to that which we will derive by using
MHV diagrams.

%As explained in \cite{Bern:1994cg}, (this will also be
%reviewed in Section 4),
In order to obtain the one-loop MHV amplitudes in
$\cN=1$ and $\cN=2$ SYM it is sufficient to compute
the contribution  $\cA^{{\cN}=1, {\rm chiral}}_{n}$
to the one-loop MHV amplitudes coming from
a single $\cN=1$ chiral multiplet.
This was calculated in \cite{Bern:1994cg},
and the result turns out to be proportional
to the Parke-Taylor MHV tree amplitude \cite{mhv}
\beq
\cA_n^{\rm tree} \ := \
{\lan i \, j \ran^4
\over
\prod_{k=1}^{n}
\lan k\, k+1 \ran
}
\ ,
\eeq
as is also the case with the one-loop 
MHV  amplitudes in $\cN=4$ SYM.
However, in contradistinction with that case,
the remaining part of the $\cN=1$ amplitudes 
depends non-trivially on the
position of the negative helicity gluons $i$ and $j$.
The result obtained in \cite{Bern:1994cg} is:
\beqa
\label{BDDKNeq1}
\cA^{{\cN}=1, {\rm chiral}}_{n}
&=&
\cA^{\rm tree}_{n}\, \cdot \,
\biggl\{ \sum_{m=i+1}^{j-1}
\sum_{s=j+1}^{i-1} b_{m,s}^{i,j}
\,
B(t_{m+1}^{[s-m]},t_m^{[s-m]}, 
t_{m+1}^{[s-m-1]},t_{s+1}^{[m-s-1]})
%\nonumber
\\
&+&
\sum_{m=i+1}^{j-1} \sum_{a\in \cD_m}
\, c^{i,j}_{m,a}
\,
{\log (t_{m+1}^{[a-m]}/t_m^{[a-m+1]})
\over
t_{m+1}^{[a-m]}-
t_m^{[a-m+1]}} \ + \
\sum_{m=j+1}^{i-1} \sum_{a\in \cC_m}
\,
c_{m,a}^{i,j}
\,
{\log (t_{a+1}^{[m-a]}/t_{a+1}^{[m-a-1]})
\over
t_{a+1}^{[m-a]}-t_{a+1}^{[m-a-1]}}
\nonumber \\
&+&{c_{i+1,i-1}^{i,j}\over
t_{i}^{[2]}} K_0(t_i^{[2]})
\, + \,
{c_{i-1,i}^{i,j}
\over
t_{i-1}^{[2]}}
K_0(t_{i-1}^{[2]})
\, + \,
{c_{j+1,j-1}^{i,j}
\over t_j^{[2]}}
K_0(t_j^{[2]})\, + \,
{c_{j-1,j}^{i,j}
\over t_{j-1}^{[2]}}
K_0(t_{j-1}^{[2]})
\biggr\}
\nonumber
\ ,
\eeqa
where $t_i^{[k]} := (p_i + p_{i+1} + \cdots + p_{i+k-1})^2$
for $k\geq 0$, and $t_i^{[k]} =t_i^{[n-k]}$ for $k <0$.
The sums in the second line of
\eqref{BDDKNeq1} cover the ranges
$\cC_m$ and $\cD_m$ defined by
\beqa
\cC_m &=&
\left\{
\begin{tabular}{ll}
$\{i,i+1, \ldots ,j-2 \}, \hskip 1.4cm   m=j+1,$
%& $m=j+1$
\cr \cr
$ \{i,i+1, \ldots ,j-1 \} , \hskip 1.4cm  
j+2 \leq m \leq i-2 , $
%&  $ j+1 <m < n ,$
\cr \cr
$\{i+1,i+2, \ldots ,j-1 \} , \hskip .7cm m =i-1,$
\end{tabular}\right.
\eeqa
and
\beqa
\cD_m &=&
\left\{
\begin{tabular}{ll}
$\{j,j+1, \ldots ,i-2 \}, \hskip 1.4cm  m=i+1,$
\cr \cr
$ \{j,j+1, \ldots ,i-1 \} , \hskip 1.4cm  i+2 
\leq m \leq j-2 , $
\cr \cr
$\{j+1,j+2, \ldots ,i-1 \} , \hskip .7cm m =j-1.$
\end{tabular}\right.
\eeqa
The coefficients
$b_{m,s}^{i,j}$ and $c_{m,a}^{i,j}$ are
\beq
\label{bdef}
b_{m, s}^{i, j} \ :=  \
-2 \, \frac{
\tr_{+} \left( \kslash_{i} \kslash_{j} \kslash_{m}
\kslash_{s} \right)
\tr_{+} \left( \kslash_{i} \kslash_{j} \kslash_{s}
\kslash_{m} \right) }
{ [ (k_i + k_j)^{2}]^2
\,
[( k_{m} + k_s)^2]^2}
\ ,
\eeq
\beq
\label{cdef}
c_{m, a}^{i, j} \ := \
\left[
{\tr_{+} ( \kslash_{m}\kslash_{a+1}\kslash_{j} \kslash_{i}   )
\over (k_{a+1} + k_m )^2}
 \, - \,
{\tr_{+} ( \kslash_{m}\kslash_{a}\kslash_{j} \kslash_{i}   )
\over (k_{a} + k_m )^2}
\right]
\,
{ {   \tr_+ ( \kslash_{i}\kslash_{j} \kslash_{m}
\qslash_{m,a} ) - \tr_+ ( \kslash_{i} \kslash_{j} 
\qslash_{m,a} \kslash_{m} )
\over [(k_i + k_j)^2]^2}}
\, ,
\eeq
where $q_{r,s}:=\sum_{l=r}^{s} k_l$.
Notice that both coefficients $b_{m, s}^{i, j}$
and $c_{m, a}^{i, j}$ are symmetric under
the exchange of $i$ and $j$.
In the case of $b$ this is evident; for $c$,
this is also manifest as it is expressed as the product of
two antisymmetric quantities.
%%%%%%%%%%%%%%%%%%%%%%%%%%%%%%%%%%%%%%%%%%%%%%%%%%%%%%%%%55%
\begin{figure} [ht]
\label{fig1}
\vspace{.2in}
\centerline {
\includegraphics[width=4in]{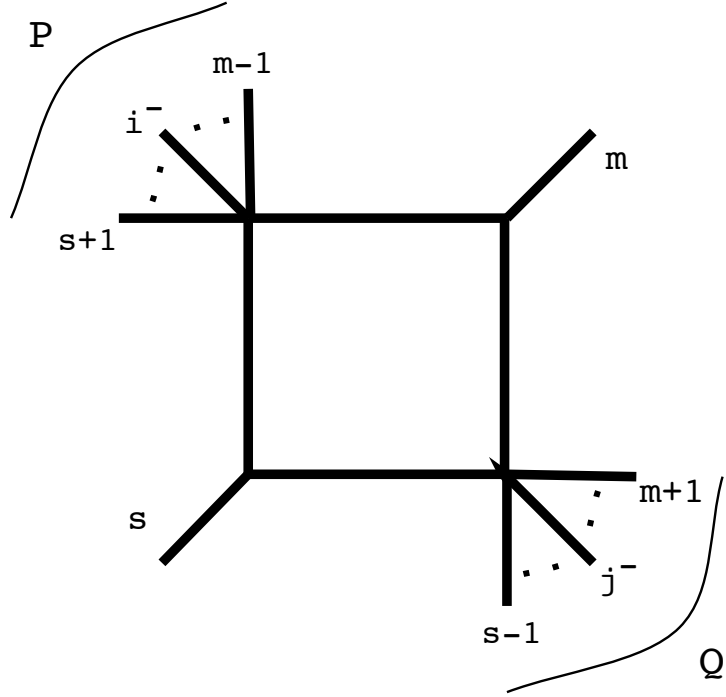}
}
\vspace{.2in}
\caption{\it
The box function $F$ of \eqref{F}, whose finite part
$B$, Eq.~\eqref{Bniceonecyril},
appears in the $\cN=1$ amplitude \eqref{BDDKNeq1}.
The two external gluons with negative helicity
are labelled by $i$ and $j$.
The legs labelled by $s$ and $m$ correspond to the null momenta
$p$ and $q$ respectively 
in the notation of \eqref{Bniceonecyril}.
Moreover, the quantities
$t_{m+1}^{[s-m]}$, $t_{m}^{[s-m]}$,
$t_{m+1}^{[s-m-1]}$, $t_{s+1}^{[m-s-1]}$
appearing in the box function $B$ in
\eqref{Neq1}
correspond to the  kinematical invariants
$t:=(Q+p)^2$, $s:=(P+p)^2$, $Q^2$, $P^2$ in the notation of
\eqref{Bniceonecyril}, with $p+q+P+Q=0$.
}
\end{figure}
\begin{figure} [ht]
\label{fig2}
\vspace{.2in}
\centerline {
\includegraphics[width=4in]{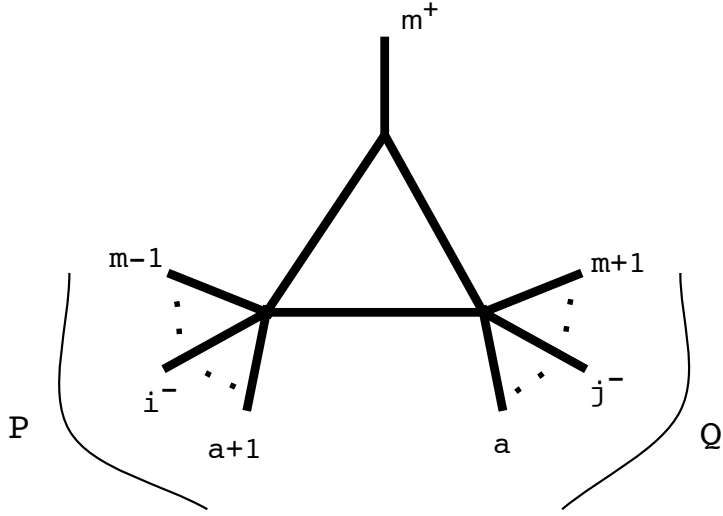}
}
\vspace{.2in}
\caption{\it
A triangle function, corresponding to the first term
$T_{\epsilon} ( p_m, q_{a+1,m-1}, q_{m+1, a})$
in the second line of \eqref{Neq1}.
$p$, $Q$ and $P$ correspond to $p_m$, $q_{m+1, a}$
and $q_{a+1, m-1}$ in the notation of Eq.~\eqref{Neq1},
where $j\in Q$, $i\in P$.
In particular, $Q^2 \to   t_{m+1}^{[a-m]}$ and
$P^2 \to t_{m}^{[a-m+1]}$.
}
\end{figure}
\begin{figure} [ht]
\label{fig3}
\vspace{.2in}
\centerline {
\includegraphics[width=4in]{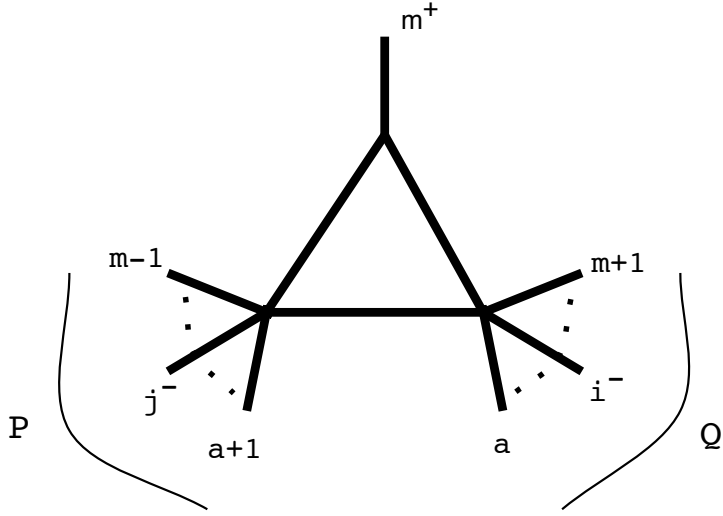}
}
\vspace{.2in}
\caption{\it
This  triangle function  corresponds to the second term
%$T_{\epsilon} ( p_m, q_{a+1,m-1}, q_{m+1, a}$
in the second line of \eqref{Neq1} -- where $i$ and $j$
are swapped.
As in Figure 2, $p$, $Q$ and $P$ correspond to
$p_m$, $q_{m+1, a}$ and $q_{a+1, m-1}$ in
the notation of Eq.~\eqref{Neq1},
where now $i\in Q$, $j\in P$.
In particular, $Q^2 \to   t_{a+1}^{[m-a]}$ and
$P^2 \to t_{a+1}^{[m-a-1]}$.}
\end{figure}
%%%%%%%%%%%%%%%%%%%%%%%%%%%%%%%%%%%%%%%%%%%%%%%%%%%%%%%%%%%%%%%%%%%%%
The function $B$ in the first line of  \eqref{BDDKNeq1}
is the ``finite'' part of the
%$\cN =4$
easy two-mass  (2me) scalar box function
$F (s,t,P^2,Q^2)$, with
\beq
\label{F}
F (s,t,P^2,Q^2) \ := \
-{1\over \epsilon^2} \Bigl[ (-s)^{-\epsilon} \, +\,
(-t)^{-\epsilon} \, -\, (-P^2)^{-\epsilon} \, -\,
(-Q^2)^{-\epsilon} \Bigr] \ + \
B (s,t,P^2,Q^2) \ .
\eeq
As in \cite{bst}, we have introduced the following 
convenient kinematical invariants: 
\beq
s\ := \ (P+p)^2 \, , \qquad t \ := \ (P+q)^2
\ , 
\eeq  
where $p$ and $q$ are null momenta, and $P$ and $Q$ are 
in general massive, with $p+q+P+Q=0$.%
\footnote{
The kinematical invariant $s=(P+p)^2$ should not be confused with
the label $s$ which is also used to label an external leg (as in Figure 1 for example). 
The correct meaning will be clear from the context.}
In \cite{bst} the following new expression for
$B$ was found:
 \beq
\label{Bniceonecyril}
 B (s,t,P^2, Q^2) \ = \
\Li2(1-aP^2)\, + \, \Li2(1-aQ^2)  \, -\,  \Li2(1-as)
\,  -\,  \Li2(1-at)
\ ,
\eeq
where
\beq
\label{aagainagain} a \ = \
\frac{P^2+Q^2-s-t}{P^2Q^2-st} 
%\ = \
%\frac{u}{P^2Q^2-st}
\ .
\eeq
%with $u:= P^2+Q^2-s-t$.
The expression \eqref{Bniceonecyril}
%found in \cite{bst}
contains one less dilogarithm and one less logarithm
than  the traditional form used by BDDK,
\beqa
 B (s,t,P^2, Q^2) & = &
\nonumber
{\rm Li}_2 \Bigl( 1 - {P^2\over s } \Bigr) \, +\, {\rm Li}_2
\Bigl( 1 - {P^2\over t } \Bigr) +
{\rm Li}_2 \Bigl( 1 - {Q^2\over
s} \Bigr) \, + \, {\rm Li}_2 \Bigl( 1 - {Q^2\over t } \Bigr)
\\ %\nonumber
&-& {\rm Li}_2 \Bigl( 1 - {P^2 Q^2\over s\, t } \Bigr)
\,
+\,
{1\over 2} \log^2 \Bigl( {s\over t} \Bigr)
\ .
\label{Bboxcsw22}
\eeqa
The agreement of \eqref{Bniceonecyril} with
\eqref{Bboxcsw22}  was discussed and proved in
Section 5 of \cite{bst}.%
\footnote{More precisely, this agreement holds only in certain kinematical
regimes e.g.~in the Euclidean region where all kinematical invariants are negative. 
More care is needed when
analytically continuing the amplitude to the physical region.  The usual prescription of
replacing a kinematical invariant $s$ by $s + i\varepsilon$ and continuing $s$ from negative
to positive values gives the correct result only for our form of the 
box function \eqref{Bniceonecyril},
whereas \eqref{Bboxcsw22} has to be amended by correction terms
\cite{binoth}.}
In Figure 1 we give a pictorial representation of
the box function  $F$ defined in \eqref{F} (with the leg labels identified
by $s \to p$, $m \to q$).

Finally, infrared divergences
are contained in the bubble functions
$K_0 (t)$, defined by
\beq
K_0 (t) \ := \
{\, {(-t)}^{-\epsilon}\over \epsilon (1-2\epsilon)}
\ .
\eeq
We  notice that in order to re-express
\eqref{BDDKNeq1} in a simpler form, it is useful 
to introduce the triangle function \cite{csw2}
\beq
\label{trian}
T(p, P, Q) \ := \
{ \log (Q^2 / P^2) \over Q^2 - P^2}
\ ,
\eeq
with $p+P+Q=0$.
A diagrammatic representation of this function
is given in Figure 2  (with $m^+ \to p$).
We also find it useful to introduce
an $\epsilon$-dependent triangle function,%
\footnote{The function $T_{\epsilon} (p, P, Q)$
defined in \eqref{epstriangle} arises naturally
in the twistor-inspired approach
which will be developed in Sections 3 and 4.}
\beq
\label{epstriangle}
T_{\epsilon} (p, P, Q) \ := \
{1\over \epsilon}
{ (-P^2)^{-\epsilon} - (-Q^2)^{-\epsilon} \over Q^2 - P^2}
\ .
\eeq
As long as $P^2$ and $Q^2$ are non-vanishing, one has
\beq
\label{T1}
\lim_{\epsilon \to 0} T_{\epsilon} (p, P, Q) \ = \
T (p, P, Q) \ , \qquad
P^2 \neq 0 \, , \, Q^2 \neq 0
\ .
\eeq
If either of the invariants vanishes, one has a different
limit. For example, if
$Q^2=0$, one has
\beq
\label{T2}
\left. T_{\epsilon} (p, P, Q)
\right|_{Q^2=0}  \ \longrightarrow \
- {1\over \epsilon} \, {(-P^2)^{-\epsilon }\over P^2}\, ,
\qquad {\epsilon \to 0}
\ .
\eeq
We will call these cases ``degenerate triangles''.

The usefulness of the  previous remark stems from
the fact that precisely the quantity
$(1 / \epsilon)\cdot \bigl[(-P^2)^{-\epsilon}/ P^2 \bigr]$
appears in the last line  of \eqref{BDDKNeq1} -- the
bubble contributions.
Therefore, these can be equivalently obtained
as  degenerate triangles, i.e.~triangles where
one of the massive legs becomes massless.

Specifically, we notice that
the four degenerate triangles (bubbles) in the last line
of \eqref{BDDKNeq1} can be precisely obtained by
including the ``missing'' index assignments
in $\cD_m$ and   $\cC_m$:
\beq
(m=i+1,\,  a=i-1) \, ,
\ \ \ (m=j-1,\,  a=j) \qquad
{\rm for} \ \cD_m \, ,
\
\eeq
which correspond to two degenerate triangles,
and
\beq
(m=j+1, \, a=j-1)\, ,
\ \ \ (m=i-1,\,  a=i) \qquad
{\rm for} \ \cC_m \, ,
\eeq
corresponding to two more degenerate triangles.

In conclusion, the previous remarks
allow us to rewrite \eqref{BDDKNeq1} in a
more  compact form as follows:
\beqa
\label{Neq1}
\cA^{{\cN}=1, {\rm chiral}}_{n}
&=&
\cA^{\rm tree}_{n}\, \cdot \,
\biggl\{ \sum_{m=i+1}^{j-1}
\sum_{s=j+1}^{i-1} b_{m,s}^{i,j}
\,
B(t_{m+1}^{[s-m]},t_m^{[s-m]},
t_{m+1}^{[s-m-1]},t_{s+1}^{[m-s-1]})
%\nonumber
\\
&+&
{1\over 1-2\epsilon}
\bigg[ \sum_{m=i+1}^{j-1} \sum_{a=j}^{i-1}
\, c^{i,j}_{m,a}
\, T_{\epsilon} ( p_m, q_{a+1,m-1}, q_{m+1, a} )
\ \, + \ \,  ( i \longleftrightarrow j )
\bigg]
\biggr\}
\nonumber
\ .
\eeqa
In the previous expression it is understood that we only keep terms that survive in
the limit $\epsilon \to 0$. This means that the factor 
$1/(1-2\epsilon) $ can be replaced by $1$ whenever the term in the sum
is finite, i.e.~whenever the triangle is non-degenerate.
However,  in the case of degenerate triangles, which contain infrared divergent
terms, we have to expand this factor to linear order in $\epsilon$. 
In the notation of \eqref{Neq1},
$q_{m+1, a}^2 =  t_{m+1}^{[a-m]}$ and
$q_{a+1, m-1}^2 =t_{m}^{[a-m+1]} $;
in Figure 2, these invariants
correspond to $Q^2$ and $P^2$ respectively,
where $j\in Q$, $i\in P$.
In the sum with $i\leftrightarrow j$, one would have
$q_{m+1, a}^2= t_{a+1}^{[m-a]}$, 
$q_{a+1, m-1 }^2=t_{a+1}^{[m-a-1]}$, corresponding
respectively to ${Q}^2$ and ${P}^2$
in Figure 3,
with $i\in {Q}$, $j\in {P}$.
It is the expression \eqref{Neq1} for the $\cN=1$
chiral multiplet
amplitude which we will derive using MHV diagrams.

%%%%%%%%%%%%%%%%%%%%%%%%%%%%%%%%%%%%%%%%%%%%%%%%%%%%%%%%%%%%
%
\section{A brief review of
loop diagrams from MHV vertices}

In this Section we briefly review the method
proposed in \cite{bst} to compute loop amplitudes in
supersymmetric gauge theories,
referring the reader to Sections 3--5 of
that paper for a more detailed discussion
of this procedure, as well as for an example of
its application to the calculation
of one-loop MHV amplitudes in $\cN=4$ SYM.

Before starting, let us first recall that
in what follows we will be dealing with
the so-called colour-stripped or partial amplitudes.
More specifically, for an $n$-particle
scattering amplitude we will compute the term
proportional to ${\rm Tr} ( T^{a_1} \cdots T^{a_n} )$,
where the $T^a$'s are the generators of the gauge group.
The full planar amplitude is then obtained by summing over
non-cyclic permutations. It is a remarkable result of BDDK that,
at one loop, non-planar amplitudes are simply obtained as a sum 
over permutations of the planar ones. This is discussed in
Section 7 of \cite{Bern:zx},
where it was also noted that this applies to a generic theory
with adjoint particles running in the loops,
such as $\cN=1,2,4$ SYM.
At the level of group theory factors, the diagrammatics for building one-loop
MHV amplitudes in the BDDK cut-constructibility approach
and in the approach discussed in this paper
(as well as in \cite{bst}, for the $\cN=4$ case)
are precisely the same. Hence, the agreement at one loop
between the two methods at planar level
in fact trivially extends to subleading corrections in $1/N$.

We now come back to the description of the approach of
\cite{bst} to loop amplitudes.
The procedure used there was:
\begin{itemize}
\item[{\bf 1.}] 
Lift the MHV tree-level scattering amplitudes to vertices, by 
continuing the internal lines off shell
using a prescription equivalent to that of CSW.
Internal lines are then connected by scalar propagators
which join particles of the same spin but opposite helicity.
\item[{\bf 2.}]
Build MHV diagrams with the required external particles
at loop level using the MHV tree-level vertices,
and sum over all independent diagrams
obtained in this fashion for a fixed ordering
of external helicity states.
\item[{\bf 3.}]
Re-express the loop integration measure in terms
of the off-shell parametrisation employed for
the loop momenta.
\item[{\bf 4.}]
Analytically continue to $4-2\epsilon$ dimensions
in order to deal with infrared divergences,
and perform  all loop integrations.
\end{itemize}
Using this method, we will show in Section 4
that combining MHV vertices into one-loop diagrams
precisely yields the results for the contribution
of the chiral multiplet to MHV amplitudes,
as obtained by BDDK using
the cut-constructibility approach.

We start off by discussing the off-shell continuation
which was used in \cite{bst},
and found to be very useful 
for calculating loop diagrams.
Consider a generic off-shell momentum vector, $L$.
On general grounds, it can be decomposed as
\cite{Bena:2004ry,dk}
\beq
\label{off}
L \ = \ l \, + \, z \eta \ ,
\eeq
where $l^2=0$, and $\eta$ is a fixed and arbitrary
null vector, $\eta^2=0$; $z$ is a real number.
Equation \eqref{off} determines $z$
as a function of $L$:
\beq
z \ = \
{L^2 \over  2 (L \eta)}
\ .
\eeq
Using spinor notation, we can write $l$ and $\eta$
as $l_{\a \da} = l_{\a} \llt_{\da}$,
$\eta_{\a \da} = \eta_{\a} \tilde{\eta}_{\da}$.
It then follows that%
\footnote{Spinor inner products
are defined as
$\lan \l \, \m \ran := \epsilon_{\a \beta} \l^{\a} \mu^{\beta}$,
$[\lt \, \tilde{\m} ]  :=
\epsilon_{\da \dot\beta} \lt^{\da} \tilde{\mu}^{\dot{\beta}}$.}
\beqa
\label{1}
l_\a & = & {L_{\a \da} \tilde{\eta}^{\da}
\over [ \llt \, \tilde{\eta}]
}
\ ,
\\
\label{2}
\llt_{\da} & = & {\eta^\a L_{\a \da} \over \lan l \, \eta
\ran} \ .
 \eeqa
We notice that  \eqref{1} and \eqref{2} coincide
with the CSW prescription proposed in\cite{csw}
to determine the spinor variables
$l$ and $\llt$ associated with the non-null,
off-shell four-vector $L$ defined in \eqref{off}.
The denominators on the right
hand sides of \eqref{1} and \eqref{2} turn out to be
irrelevant for our applications, since the expressions
we will be dealing with are homogeneous in the spinor
variables $l_{\alpha}$; hence we will discard them.

To proceed further, we need to re-express the usual integration
measure $d^4L$ over the loop momentum $L$ in terms
of the new variables $l$ and $z$ introduced previously.
We found that
\footnote{The $i \varepsilon$ prescription
in the left and right hand sides of
\eqref{mmm} was understood in \cite{bst}, and,
as stressed in \cite{Lance,Bena:2004ry},
it is essential in order to correctly perform
loop integrations.}
\beq
\label{mmm}
{d^4 L \over
L^2 + i \varepsilon} \ = \
{d\cN (l) \over 4i}
\, {dz \over z + i\varepsilon} \ ,
\eeq
where we have introduced the Nair measure  \cite{Nair}
\beq
 d\cN (l) := \, \lan l
\, \, dl \ran \, d^2 \tilde{l} \ - \ [ \tilde{l} \, \, d
\tilde{l}] \, d^2 l \ .
\eeq
Eq.~\eqref{mmm} is key to the procedure.
It is important to notice that
the product of the measure factor with
a scalar propagator $d^4 L / (L^2 + i \varepsilon)  $
in \eqref{mmm} is independent of the reference vector $\eta$.
In \cite{Nair}, it was noticed that the Lorentz invariant
phase space measure for a massless particle can be expressed
precisely in terms of the Nair measure:
\beq
\label{nairmeas}
 d^4l \, \delta^{(+)} (l^2)\ =  \
{d\cN (l)\over 4i} \ ,
%{i\over 4} \ d\cN (l)  \ ,
\eeq
where, as before, we write the null vector $l$ as $l_{\a \da} =
l_{\a} \tilde{l}_{\da}$, and in Minkowski space we identify
$\tilde{l} = l^{\ast}$.

Next, we observe that in computing one-loop diagrams,
the four-dimensional integration measure which appears is
\beq
\label{dem}
d\cM \ := \ {d^4 L_1 \over L_1^2} {d^4 L_2 \over L_2^2}
\,
\delta^{(4)} (L_2 - L_1 + P_L)
\ ,
\eeq
where $L_1$ and $L_2$ are loop momenta, and $P_L$ is the
external momentum flowing outside the loop%
\footnote{In our conventions, all external momenta are
outgoing.}
so that $L_2 - L_1 + P_L=0$.
Next we express  $L_1$ and $L_2$ as in
\eqref{off},
\beq
\label{ells}
 L_{i; \a, \da} \ =
\ l_{i \a} \llt_{i \da}  \, + \,
z_i \, \eta_{\a}\tilde{\eta}_{\da} \ ,
\qquad i=1,2 \ .
 \eeq
Using \eqref{ells},
we  rewrite the argument of the delta function as
\beq
L_2 - L_1 + P_L
= l_2 - l_1 + P_{L; z} \ ,
\eeq
where we have defined
\beq
\label{plz}
P_{L;z} := P_L - z \eta
\ ,
\eeq
and
\beq
z \ := \ z_1\, - \, z_2
\ .
\eeq
Notice that we use the same $\eta$ for both the momenta
$L_1$ and $L_2$.
Using \eqref{ells}, we can finally recast \eqref{dem} as
\cite{bst}
\beq
\label{gcar}
d\cM \ = \
%-4  \, 
{dz_1 \over z_1}\, { dz_2 \over z_2}
\
d{\rm LIPS} (l_2 , - l_1;P_{L;z})
\ ,
\eeq
where
\beq
\label{LIPS}
d{\rm LIPS} (l_2 , - l_1;P_{L;z}) \ := \
d^4 l_1 \, \delta^{(+)} (l_1^2) \
d^4 l_2 \, \delta^{(+)} (l_2^2 )\
\delta^{(4)} (l_2 - l_1 + P_{L;z})
\
\eeq
is the two-particle Lorentz invariant phase
space  (LIPS) measure.
The integration measure $d\cM$ as it is expressed
on the right hand side of \eqref{gcar}
can now be immediately dimensionally regularised;
this is accomplished by simply replacing
the four-dimensional LIPS measure by
its continuation to $D=4-2\epsilon$ dimensions,
\beq
\label{LIPSD}
d^D{\rm LIPS} (l_2 , - l_1;P_{L;z}) \ := \
d^D l_1 \, \delta^{(+)} (l_1^2) \
d^D l_2 \, \delta^{(+)} (l_2^2 )\
\delta^{(D)} (l_2 - l_1 + P_{L;z})
\ .
\eeq
Eq.~\eqref{gcar} was  one of the
key results of  \cite{bst}.
It gives a decomposition of the original
integration measure into a $D$-dimensional
phase space measure and a
dispersive measure.
According to Cutkosky's cutting rules
\cite {Cutkosky:1960sp}, the LIPS measure computes 
the discontinuity of a Feynman diagram across
its branch cuts. Which discontinuity is evaluated
is determined by the argument of the delta function
appearing in the LIPS measure;
in \eqref{gcar} this is $P_{L;z}$ (defined in
\eqref{plz}).
Notice that $P_{L;z}$ always contains a term
proportional to the reference vector $\eta$,
as prescribed by  \eqref{plz}.
Finally, discontinuities are integrated using the
dispersive measure in \eqref{gcar},
thereby reconstructing the full amplitude.

As a last remark, notice that,
in contradistinction with the cut-constructibility
approach of BDDK, here we sum over all the cuts --
each of which is integrated with the appropriate
dispersive measure.

%%%%%%%%%%%%%%%%%%%%%%%%%%%%%%%%%%%%%%%%%%%%%%%%%%%%%%%%%%%%

\section{MHV one-loop amplitudes in $\cN=1$ SYM 
from MHV vertices}

In the last section we reviewed how MHV vertices can be sewn
together into one-loop diagrams, 
and how a particular decomposition
of the loop momentum measure leads to a representation of the
amplitudes strikingly similar to traditional dispersion formulae.
This method was tested successfully in 
\cite{bst} for the case of
MHV one-loop amplitudes in $\cN=4$ SYM.
In the following we will apply the same philosophy to amplitudes
in $\cN=1$ SYM, in particular to the infinite sequence of
MHV one-loop amplitudes, which was obtained using
the cut-constructibility approach \cite{Bern:1994cg},
and whose twistor space picture has
more recently been analysed in \cite{csw2}.

Similarly to the $\cN=4$ case, the one-loop
amplitude has an overall factor
proportional to the MHV tree-level amplitude,
but, as opposed to the $\cN=4$ case,
the remaining one-loop factor depends non-trivially
on the positions $i$ and $j$ of the two
external negative helicity gluons.
This is due to the fact that
a different set of fields is allowed
to propagate in the loops.

The MHV Feynman diagrams contributing to
MHV one-loop amplitudes consist of two
MHV vertices connected by two
off-shell scalar propagators. If both
negative helicity gluons are on one
MHV vertex, only gluons of a particular
helicity can propagate in the loop. This is 
independent of the number of supersymmetries.
On the other hand, for diagrams with one negative
helicity gluon on one MHV vertex and the other
negative helicity gluon on the other
MHV vertex, all components of the
supersymmetric multiplet propagate in the
loop. In the case of $\cN=4$ SYM
this corresponds to helicities $h=-1,-1/2,0,1/2,1$
with multiplicities $1,4,6,4,1$, respectively; 
for the $\cN=1$ vector multiplet
the multiplicities are $1,1,0,1,1$.
Hence, we can obtain the $\cN=1$ amplitude
by simply taking the $\cN=4$ amplitude
and subtracting three times the contribution
of an $\cN=1$ chiral multiplet,
which has multiplicities $0,1,2,1,0$.%
\footnote{We can also obtain the $\cN=2$ amplitude 
in a completely similar way.}

This supersymmetric decomposition of
general one-loop amplitudes is useful
as it splits the calculation into pieces of
increasing difficulty, and allows one to reduce
a one-loop diagram with gluons
circulating in the loop to a combination
of an $\cN=4$ vector amplitude,
an  $\cN=1$ chiral amplitude and finally
a non-supersymmetric amplitude with
a scalar field running in the loop.

In our case, the supersymmetric
decomposition takes the form
\begin{equation}
\label{susydec}
\cA_n^{\cN=1, {\rm vector}} \ = \
\cA_n^{\cN=4}\, - \, 
3 \, \cA_n^{\cN=1,{\rm chiral}}
\ ,
\end{equation}
where $n$ denotes the number of external lines.
Since the $\cN=4$ contribution is known,  one needs to
determine $\cA_n^{\cN=1,{\rm chiral}}$ using MHV diagrams.
To be precise, we are only addressing the computation of
the planar part of the amplitudes.
%in 't Hooft's sense.
However this is sufficient, since at one-loop level
the non-planar partial amplitudes are obtained
as appropriate sums of permutations of
the planar partial amplitudes \cite{Bern:zx},
as discussed at the beginning of Section 3.
%%%%%%%%%%%%%%%%%%%%%%%%%%%%%%%%%%%%%%%%%%%%%%%%%%%%%%%%%%
\begin{figure} [ht]
\label{fig4}
\vspace{.2in}
\centerline {
\includegraphics[width=4in]{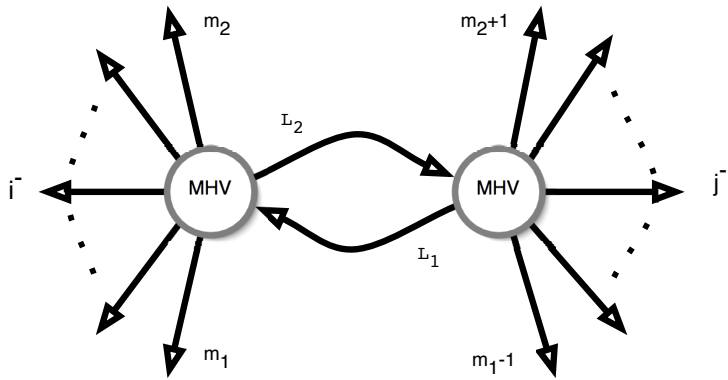}
}
\vspace{.2in}
\caption{\it A one-loop MHV  diagram,
computed in \ref{loopint} using MHV amplitudes 
as interaction vertices,
with the CSW off-shell prescription.
The two external gluons with negative helicity
are labelled by $i$ and $j$.}
\end{figure}
%%%%%%%%%%%%%%%%%%%%%%%%%%%%%%%%%%%%%%%%%%%%%%%%%%%%%%%%%%%%%

Therefore our task consists of
\begin{itemize}
\item[{\bf 1.}]
Evaluating the class of diagrams
where we allow all the helicity states
of a chiral multiplet,
\beq
\label{rangeh}
h \in \{-1/2,0,0,1/2\}
\ ,
\eeq
to run in the loop.
We depict the prototype of such diagrams in
Figure 4.
\item[{\bf 2.}]
Summing over all diagrams such that each of
the two MHV vertices always has one external
gluon of negative helicity. Assigning
$i^-$ to the left and $j^-$ to the right,
the summation range of $m_1$ and $m_2$ is determined to be:
\beq
\label{range}
j +1\leq m_1 \leq i \ , \qquad
i \leq m_2 \leq j-1 \ .
\eeq
\end{itemize}
Therefore we get
\beqa
\label{loopint}
\cA_n^{\cN=1,{\rm chiral}} & = & \sum_{m_1,m_2,h}
\int\!d\cM \,
%\frac{d^4L_1}{L_1^2 + i \epsilon}
%\frac{d^4L_2}{L_2^2 + i \epsilon}
\cA(-l_1,m_1,\ldots,i^{-},\ldots,m_2,l_2) \nonumber \\
& & \,\,\,\,\,\,\,\,\,\,\,
\cdot \,
\cA(-l_2,m_2+1,\ldots,j^{-},
\ldots,m_1-1,l_1)
\,\, ,
\eeqa
where the summation ranges of $h$, $m_1$ and $m_2$ are
given in \eqref{rangeh}, \eqref{range}.
Notice that, in order to compute the loop amplitude 
\eqref{loopint}, we make use of the integration measure 
$d\cM$, given in \eqref{gcar},
which was found  in \cite{bst}.

After some spinor algebra and after performing the sum
over the helicities $h$, the integrand of
(\ref{loopint}) becomes
\beq
\label{integrand}
-i\cA_n^{\rm tree} \, \cdot \,
\frac{\langle m_2\,  (m_2 \! + \! 1)
\rangle \, \langle (m_1 \! - \! 1) \, m_1 \rangle
\,
\langle i\,  l_1 \rangle
\,
\langle j \, l_1 \rangle \langle i \, l_2 \rangle
\langle j \, l_2 \rangle}{\langle i\,  j\rangle^2
\,
\langle m_1 \, l_1 \rangle
\,
\langle (m_1 \! - \! 1) \, l_1 \rangle
\,
\langle m_2 \, l_2 \rangle
\langle (m_2 \! + \! 1) \, l_2\rangle}
\,\, .
\eeq
The focus of the remainder of this section
will be to evaluate  the
integral in \eqref{loopint} explicitly.
Since $-i \cA_n^{\rm tree}$ factors out
completely, we will drop it
and only reinstate it at the very end
of the calculation.

The integrand (without this factor)
can be rewritten in terms of dot products of
momentum vectors,
\beq
\label{redintegrand}
{\mathcal I} \ = \
\frac{\mathcal N}{(i \cdot j)^{2}
\left( m_1 \cdot l_1 \right)
\left( (m_1 \!- \! 1 ) \cdot l_1 \right)
\left( m_2 \cdot l_2 \right)
\left( (m_2 \! + \! 1) \cdot l_2 \right) } \,\, ,
\eeq
with
\beq
\label{numerator}
{\mathcal N} = \tr_+ \left( \lslash_1 \kslash_{m_1-1} \kslash_{m_1}
\lslash_1 \kslash_j \kslash_i \right)
\tr_+ \left( \lslash_2 \kslash_{m_2} \kslash_{m_2+1}
\lslash_2 \kslash_j \kslash_i \right)
\,\, .
\eeq
$\cN$ is a product of Dirac traces,
where the $\tr_+$ symbol indicates that the
projector $(1+\gamma^{5})/2$ has been inserted.

Next, notice that each of these Dirac
traces involving six momenta 
can be expressed in terms of simpler
Dirac traces involving only four momenta.
For the first factor
of (\ref{numerator}) we find
\beq
\label{6to4trace}
\tr_+ \left( \lslash_1 \kslash_{m_1-1} \kslash_{m_1}
\lslash_1 \kslash_j \kslash_i \right) \, = \,
2 (m_1 \cdot l_1) \tr_+  \left( \kslash_{i}
\kslash_{j} \kslash_{m_1-1}
\lslash_1 \right) \, - \,
2 ((m_1 \!- \!1) \cdot l_1) \tr_+
\left( \kslash_{i} \kslash_{j} \kslash_{m_1}
\lslash_1 \right) \, ,
\eeq
where
\beq
\label{4trace}
\tr_+ \left( \kslash_{a} \kslash_{b} \kslash_{c}
\kslash_d \right) = 2 \bigl[ (a \cdot b) (c \cdot d)-
(a \cdot c) (b \cdot d) +
(a \cdot d) (b \cdot c)\bigr] - 2i  \varepsilon(a,b,c,d) \, .
\eeq
The second factor in \eqref{numerator} takes a similar form.
Consequently, the integrand
becomes a sum of four terms, one of which is
\beq
\label{partintegrand}
\frac{\tr_{+} \left( \kslash_{i} \kslash_{j} \kslash_{m_{1}}
\lslash_1 \right)   \tr_{+} \left( \kslash_{i}
\kslash_{j} \kslash_{m_{2}}
\lslash_2 \right) }{(i \cdot j)^{2}
\left( m_{1} \cdot l_{1}\right) \left( m_{2} \cdot l_{2}\right)}
\, .
\eeq
The other three terms are obtained by replacing
$m_{1}$ with $m_{1}-1$ and/or
$m_{2}$ with $m_{2}+1$ in \eqref{partintegrand} and come with alternating signs.
Note that the original expression \eqref{integrand}
is symmetric in $i$, and $j$, 
although when we make use of the decomposition
\eqref{partintegrand} this symmetry
is no longer manifest. We will
symmetrize over $i$ and $j$ at the end of the calculation in order to make this
exchange symmetry manifest in the final expression.

In the next step we have to perform the
phase space integration,  which is equivalent
to the calculation of a unitarity cut with momentum
$P_{L;z}=\sum_{l=m_{1}}^{m_{2}}
k_{l} - z \eta$ flowing through the cut.
Note that, as explained in Section 3, the momentum is
shifted by a term proportional to the reference
momentum $\eta$.
The term $(l_1 \cdot m_1) (l_2 \cdot m_2)$ in the
denominator in  \eqref{partintegrand} corresponds
to two propagators, hence the denominator by itself
corresponds to a cut box diagram.
However, the numerator of \eqref{partintegrand}
depends non-trivially on the loop momentum,
so that in fact  \eqref{partintegrand} corresponds
to a tensor box diagram,  not simply a scalar box diagram.
Using the Passarino-Veltman method  \cite{pv},
we can reduce the expression \eqref{partintegrand},
integrated with the LIPS measure,  to a sum of cuts
of scalar box diagrams, scalar and vector triangle diagrams,
and scalar bubble diagrams.
This procedure is somewhat technical, and details
are collected in Appendix A.
Luckily, the final result takes a less intimidating
form  than the intermediate expressions.
We will now present the result of
these calculations after the LIPS integration.

We first observe that loop integrations
are performed in $4-2\epsilon$ dimensions.
It turns out that singular $1 / \epsilon$ terms
appearing at intermediate steps of the phase space integration
cancel out completely.
Notice that this does not mean that the final result will be
free of infrared divergences.
In fact the dispersion integral can and does give rise to
$1/ \epsilon$ divergent terms but there cannot be any
$1 / \epsilon^2$ terms, as expected 
for the contribution of
a chiral multiplet \cite{Bern:1994cg}.
The $1 / \epsilon$ divergences 
in the scattering amplitude correspond to
the bubble contributions in \eqref{BDDKNeq1},
or degenerate triangles contributions in \eqref{Neq1},
as explained in Section 2.
In Appendix A we show that the finite terms of the phase 
space integral combine into the following simple expression:
\beq
\label{Neq1cut}
\hat{{\mathcal C}}  \ = \
{\mathcal C}(m_{1}-1,m_{2})-{\mathcal C}(m_{1},m_{2})+
{\mathcal C}(m_{1},m_{2}+1)-{\mathcal C}(m_{1}-1,m_{2}+1)
\ ,
\eeq
with%
\footnote{In \eqref{partNeq1cut} we omit an overall, 
finite numerical factor
that depends on $\epsilon$. This factor, 
which can be read off from \eqref{ubi}, 
is irrelevant for our discussion.}
\beqa
\label{partNeq1cut}
{\mathcal C}(m_{1},m_{2}) & = & \frac{2 \pi}{1-2 \epsilon}
\frac{ (P_{L;z}^2)^{-\epsilon}}{(i \cdot j)^{2} (m_{1} \cdot m_{2})}
\left[
\frac{{\mathcal T}(m_{1},m_{2},P_{L;z})}
{\left( m_{1} \cdot P_{L;z}\right)}
\,  +  \,
\frac{{\mathcal T}(m_{2},m_{1},P_{L;z})}
{\left( m_{2} \cdot P_{L;z}\right)}
\right] \nonumber \\[12pt]
& -&
\frac{2 \pi {\mathcal T}(m_{1},m_{2},m_{2})}
{(i \cdot j)^{2}(m_{1} \cdot m_{2})^{2}}
\, (P_{L;z}^2)^{-\epsilon}\, \log \left( 1 -
%\frac{(m_{1} \cdot m_{2})}{2 N(P_{L;z})}
a_z \, P^{2}_{L;z}\right)
\ ,
\eeqa
where
\beqa
\label{bla}
{\mathcal T}(m_{1},m_{2},P) & := &
\tr_{+} \left( \kslash_{i} \kslash_{j} \kslash_{m_{1}}
\Pslash \right)
\,
\tr_{+} \left( \kslash_{i} \kslash_{j} \kslash_{m_{2}}
\kslash_{m_{1}} \right) \,\, ,
\nonumber \\ \cr
a_z &:=& \frac{m_{1} \cdot m_{2}}{ N(P_{L;z})}
\ ,
\eeqa
and
\beq
\label{N}
N(P) \ := \ (m_{1} \cdot m_{2}) P^{2} - 2 (m_{1} \cdot P)
(m_{2} \cdot P) \,\, .
\eeq
A closer inspection of \eqref{partNeq1cut}
reveals that the first line of that expression
corresponds to two cuts of scalar triangle integrals,
up to an $\epsilon$-dependent
factor and the explicit $z$-dependence
of the two numerators. 
The second line is a term familiar from \cite{bst},
corresponding to the $P_{L;z}^2$-cut
of the finite part $B$ of a  scalar box function,
defined in \eqref{Bniceonecyril} (see also \eqref{F}).
The full result for the one-loop MHV amplitudes is obtained
by summing over all possible MHV diagrams,
as specified in
\eqref{loopint} and \eqref{rangeh}, \eqref{range}.

We begin our analysis by focusing on the box function
contributions in \eqref{partNeq1cut},
and notice the following important facts:
\begin{itemize}
\item[{\bf 1.}]
By taking into account the four terms in
\eqref{Neq1cut} and summing over Feynman diagrams,
we see each fixed finite box function $B$
appears in exactly four phase space integrals,
one for each of its  possible cuts,
in complete similarity with \cite{bst}.
It was shown in Section 5 of that paper
that the corresponding dispersion integration
over $z$ will then yield the finite $B$ part of the
scalar box functions $F$.
It was also noted in \cite{bst} that one
can make a particular gauge choice for $\eta$
such that the $z$-dependence in $N$ disappears.
This happens when $\eta$ is chosen to be equal to one
of the massless external legs of the
box function. The question of gauge invariance is discussed in
Appendix C.
\item[{\bf 2.}]
The coefficient multiplying the finite box function
is precisely equal to $b_{m_1, m_2}^{i, j}$
defined in \eqref{bdef}.
\item[{\bf 3.}]
Finally, the  functions $B$ generated by
summing over all MHV Feynman diagrams with the range
dictated by \eqref{range} are precisely those included
in the double sum for the finite box functions
in the first line of \eqref{BDDKNeq1} (or \eqref{Neq1})
upon identifying  $m_1$ and $m_2$ with $s$ and $m $.
To be precise, \eqref{range} includes the case where
the indices $s$ and/or $m$ (in the notation of
\eqref{BDDKNeq1} and \eqref{Neq1})
are equal to either $i$ or $j$; but for any of
these choices, it is immediate to check that
the corresponding coefficient  $b_{m, s}^{i, j}$
vanishes.
\end{itemize}
This settles the agreement between
the result of our computation with MHV vertices
and \eqref{Neq1} for the part corresponding to
the box functions.
Next we have to collect the cuts contributing to
particular triangles, and show that the
$z$-integration  reproduces the expected triangle
functions from \eqref{Neq1},
each with the correct coefficient.

To this end, we notice that for each fixed triangle
function $T(p, P, Q)$, exactly four phase space integrals appear,
two for each of the two possible cuts of the function.
Moreover, a gauge invariance similar to that
of the box functions also exists for triangle
cuts (see Appendix C), so that we can choose $\eta$ in a way
that the ${\mathcal T}$ numerators 
in \eqref{partNeq1cut}
become independent of $z$. A particularly convenient
choice is $\eta = k_{i}$, since it can
be kept fixed for all possible cuts.
Choosing this gauge, we see that a sum, ${\sf T}$, of
terms proportional to cut-triangles is generated
from  \eqref{Neq1cut}
(up to a common normalisation):
\beq
\label{popp}
{\sf T} \ := \ T_A \, + \,  T_B \, + \,  T_C \, + \,  T_D \,
\  ,
\eeq
where
\beqa
\label{Michelangeli}
T_A & := &
\left[ {\cS (i, j, m_1, m_2) \over (m_1 \cdot m_2 )}
\,  -\,
{\cS ( i , j , m_1 -1 , m_2) \over (( m_1 - 1)\cdot m_2)}
\right] \,
\cS (i, j, m_2 , P_L ) \,
\Delta_A
\ ,
\\ \nonumber \cr
T_B &:= &
 \left[ {\cS (i, j, m_2, m_1) \over (m_1 \cdot m_2 )}
 \, -\,
{\cS ( i , j , m_2 +1 , m_1) \over (( m_2 + 1)\cdot m_1)}
\right] \,
\cS (i, j, m_1 , P_L ) \,
\Delta_B
\ ,
\\ \nonumber \cr
T_C &:= &
\left[ {\cS (i, j, m_2+1 , m_1-1) 
\over ((m_2+1) \cdot (m_1-1) )}
 \, -\,
{\cS ( i , j , m_2  , m_1-1 ) \over (m_2 \cdot (m_1-1))}
\right] \,
\cS (i, j, m_1-1 , P_L ) \,
\Delta_C
\ ,
\\ \nonumber \cr
T_D  &:= &
\left[ {\cS (i, j, m_1-1, m_2+1) \over ((m_1-1) \cdot (m_2+1) )}
 \, - \,
{\cS ( i , j , m_1 , m_2+1) \over ( m_1 \cdot (m_2 +1))}
\right]\,
\cS (i, j, m_2+1 , P_L ) \,
\Delta_D
\ .
\eeqa
Here we have defined
\beq
\cS (a,b,c,d) \:= \
\tr_{+} \left( \kslash_{a} \kslash_{b} \kslash_{c}
\kslash_{d}
 \right)
\ ,
\eeq
and $\Delta_I$, $I = A, \ldots , D$,
are the following cut-triangles, all in the
$P_{L;z}$-cut :
\beqa
\label{listofct}
\Delta_A &:=& {1\over (m_2 \cdot P_{L;z} ) }
\ = \ Q^{2}\mbox{{\rm -cut}} \ {\rm of}
\ \ -T\big(m_2, P_{L;z} - m_2, -P_{L;z}\big)
\ ,
\\ \nonumber %\cr
\Delta_B &:=& {1\over (m_1 \cdot P_{L;z} )}
\ = \ P^{2}\mbox{{\rm -cut}} \ {\rm of}
\ \ -T\big(m_1,- P_{L;z}, P_{L;z}-m_1\big)\ ,
\\ \nonumber %\cr
\Delta_C &:=& {1\over ((m_1 - 1)  \cdot P_{L;z} )}
\ = \ Q^{2}\mbox{{\rm -cut}} \ {\rm of}
\ \ T\big( m_1 -1,- P_{L;z} - (m_1 -1), P_{L;z}\big)
\ ,
\\ \nonumber %\cr
\Delta_D &:=& {1\over ((m_2+ 1)  \cdot P_{L;z} )}
\ = \ P^{2}\mbox{{\rm -cut}} \ {\rm of}
\ \ T\big( m_2 + 1 , P_{L;z} , -P_{L;z}-(m_2 + 1)\big)
\ .
\eeqa

%%%%%%%%%%%%%%%%%%%%%%%%%%%%%%%%%%%%%%%%%%%%%%%%%%%%%%%%%%%%55
\begin{figure} [ht]
\label{fig5}
\vspace{.2in}
\centerline {
\includegraphics[width=4in]{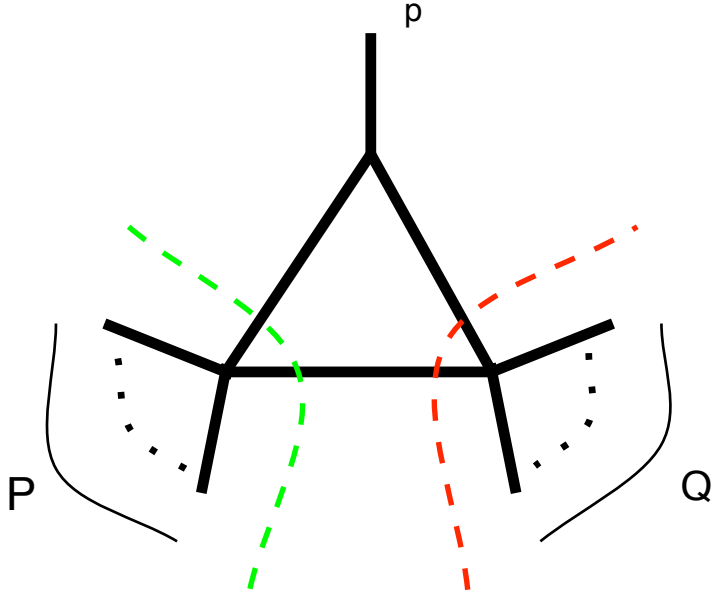}
}
\vspace{.2in}
\caption{\it A triangle function with massive
legs labelled by $P$ and $Q$, and massless leg $p$.
This function is reconstructed by summing two
dispersion integrals, corresponding to the 
$P^2_z$- and $Q^2_z$-cut.
}
\end{figure}
\begin{figure} [ht]
\label{fig6}
\vspace{.2in}
\centerline {
\includegraphics[width=4in]{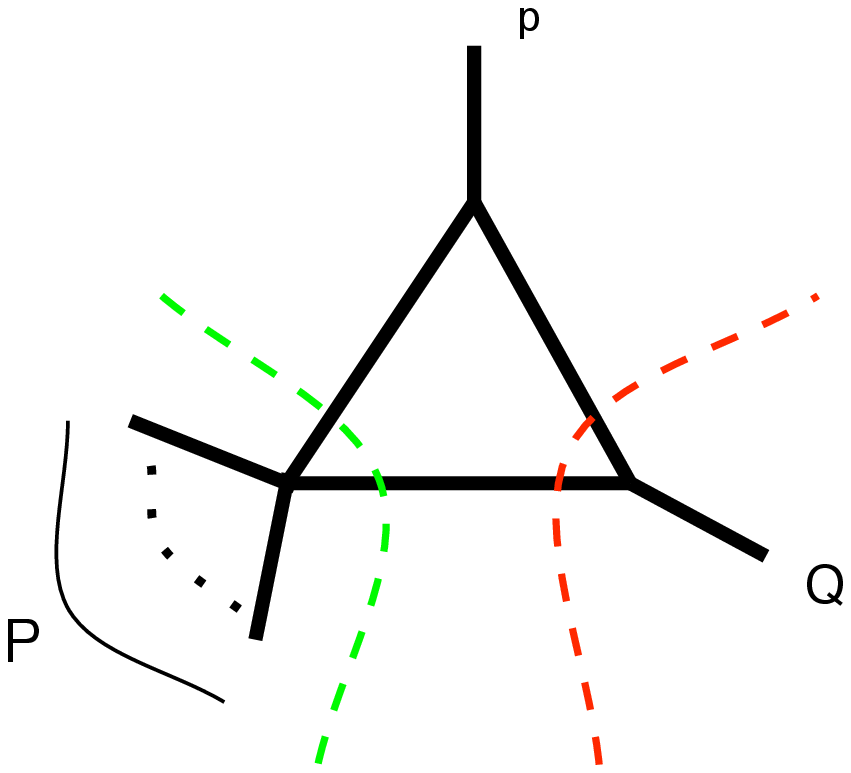}
}
\vspace{.2in}
\caption{\it
A degenerate triangle function. Here the leg labelled by $P$ is
still massive, but that labelled by $Q$ becomes massless.
This function is also reconstructed by summing over two
dispersion integrals, corresponding to the 
$P^2_z$- and $Q^2_z$-cut.
}
\end{figure}
%%%%%%%%%%%%%%%%%%%%%%%%%%%%%%%%%%%%%%%%%%%%%%%%%%%%%%%%%%%%%

Next, we notice that the prefactors multiplying
$\D_B$, $\D_C$ become the same, up to a minus sign,
upon shifting $m_1-1 \to m_1$ in the second prefactor;
and  so do the prefactors of $\D_A$, $\D_D$
upon shifting $m_2 \to m_2 + 1$.
Doing this, $-\D_B$ and the shifted  $\D_C$ become
the two cuts of the same triangle function
$T (m_1,  -P_{L;z} , P_{L;z}-m_1 )$, and similarly,
 $-\D_A$ and  $\D_D$ give the two cuts of the function
$T (m_2,  P_{L;z}-m_2, -P_{L;z}  )$.
Furthermore,
in Appendix C we will show that summing the
two dispersion integrals of the two different cuts
of a triangle indeed generates the triangle function
-- in fact this procedure gives a novel way
of obtaining the triangle functions.%
\footnote{A remark is in order here.
In our procedure the momentum appearing
in each of the possible cuts is always shifted
by an amount proportional to $z \eta$;
the triangle is then reproduced by performing
the appropriate dispersion integrals.
Because of the above mentioned shift,
we produce a non-vanishing cut (with shifted momentum)
even when the cut includes only one external (massless) leg,
say $\tilde{k}$, as the momentum flowing in the cut is
effectively $\tilde{k}_z= \tilde{k}- z \eta$,
so that  $\tilde{k}_z^2 \neq 0$.
}
Specifically, the result derived in Appendix C 
is 
\beq
\label{dfd}
\int\! {dz \over z} \left[
{(P_{z}^{2})^{-\epsilon}\over ( P_{z}p) } \, + \,
{(Q_{z}^{2})^{-\epsilon}\over (Q_{z}p)}
\right]
 \ = \ 
2 \bigl[\pi \epsilon \csc (\pi \epsilon) \bigr]
\, T_{\epsilon}(p, P, Q) 
\ , 
\eeq
where the $\epsilon$-dependent triangle function 
$ T_{\epsilon}(p, P, Q) $ (with $p+P+Q=0$)
was introduced in  \eqref{epstriangle} and gives, 
as $\epsilon \to 0$, the triangle function 
\eqref{trian} (as well as the bubbles when either
$P^2$ or $Q^2$ vanish).
The result \eqref{dfd} holds for a generic
choice of the reference vector $\eta$, 
see \eqref{bpbq}-\eqref{finaltriang}. 
We give a pictorial representation of 
the non-degenerate and degenerate triangle functions 
in Figures 5 and 6, respectively.

At this point, it should be noticed  that 
for a  gauge choice different from 
$\eta = k_i$ adopted so far, the 
numerators $\cT$ in \eqref{partNeq1cut} do acquire 
an $\eta$-dependence.
This gauge dependence should not be present 
in the final result for the scattering amplitude. 
Indeed, it is easy to check that, thanks to 
\eqref{bpbq}, the coefficient of the
$\eta$-dependent terms actually vanishes.

Using now \eqref{popp}-\eqref{dfd} and collecting 
terms as specified above, 
we see that the generic term produced by this procedure
takes the form
\beq
\label{thisterm}
\left[
{\cS ( i , j , a  , p_m) \over (k_a \cdot p_m )}
 \, - \,
{\cS ( i , j , a+1  , p_m) \over (k_{a+1} \cdot p_m )}
\right]
\,
\cS ( i , j , p_m , Q) \
T(p_m, P, Q)
\ ,
\eeq
with
$P=q_{a+1, m-1}$ and $Q=q_{m+1 , a}$.

Finally, we implement the symmetrization
of the indices $i$, $j$, as explained earlier,
and convert \eqref{thisterm} into
\beq
c_{m, a}^{i, j} \, T(p_m, P, Q)
\ ,
\eeq
where the coefficient $c_{m, a}^{i, j}$ is%
\footnote{In writing \eqref{art},
we make also use of the fact that
$\cS( i, j, q_{m-1,a}, p_m)
= \cS( i, j, q_{m,a}, p_m)$.}
\beq
\label{art}
c_{m, a}^{i, j} \ := \
{1\over 2}  \left[
{\cS ( i , j , a+1  , p_m) \over (k_{a+1} \cdot p_m )}
 \, - \,
{\cS ( i , j , a  , p_m) \over (k_{a} \cdot p_m )}
\right]
\,
{ {\cS (i, j, p_m, q_{m,a}) - \cS( i, j, q_{m,a}, p_m)
\over [(k_i + k_j)^2]^2}}
\ ,
\eeq
which  coincides with the definition of
$c_{m, a}^{i, j}$  given in \eqref{cdef}.
Finally, it is easy to see that in summing over the range
given by \eqref{range}, we produce exactly 
all the triangle functions  appearing in
the second line of \eqref{Neq1}. 
It is also important to notice that the bubbles, which
appear in the last line of \eqref{BDDKNeq1},
are actually obtained as particular cases of triangle functions 
where one of the massive legs becomes massless,
as observed at the end of Section 2.

In conclusion, we have seen that all the terms 
in \eqref{Neq1}, i.e.~finite box contributions
and triangle contributions -- 
which include the bubbles as special
(degenerate) cases -- are precisely reproduced
in our diagrammatic approach.

%%%%%%%%%%%%%%%%%%%%%%%%%%%%%%%%%%%%%%%%%%%%%%%%%%%%%%%%%%%

%%%%%%%%%%%%%%%%%%%%%%%%%%%%%%%%%%%%%%%%%%%%%%%%%%%%%%%%%%%

\section*{Acknowledgements}

It is a pleasure to thank
Lance Dixon, Valya Khoze, David Kosower, Marco Matone
and Sanjaye Ramgoolam for interesting conversations.
GT acknowledges the support of PPARC.

\newpage

%%%%%%%%%%%%%%%%%%%%%%%%%%%%%%%%%%%%%%%%%%%%%%%%%%%%%%%%%%%%
%%%%%%%%%%%%%%%%%%%%%%%%%%%%%%%%%%%%%%%%%%%%%%%%%%%%%%%%%%%%
\startappendix

\Appendix{Passarino-Veltman reduction}
In Section 4 we saw that a typical term in the
$\mathcal{N}=1$ amplitude is the dispersion integral
of the following phase space integral:
\begin{eqnarray}
\label{partintegral}
\cC (m_1 , m_2)  \ := \
\int\! d{\rm LIPS}(l_2, -l_1; P_{L;z}) \
%\frac{d^{4}L_{1}}{L_{1}^{2}}\frac{d^4 L_2}{L_2^2}
%{\delta}^{(4)}(L_2-L_1+P_L)
\frac{{\tr}_{+}(\kslash_i \kslash_j
\kslash_{m_1} \lslash_1){\tr}_{+}(\kslash_{i}\kslash_{j}
\kslash_{m_2}\lslash_{2})}{(i\cdot j)^2(m_1\cdot l_1)
(m_2\cdot l_2)}
\ .
\end{eqnarray}
The full amplitude is then obtained by adding
the dispersion integrals of three more terms
similar to  \eqref{partintegral} but
with $m_1$ replaced by $m_1-1$ and/or
$m _2$ replaced by $m_2+1$.
The goal of this Appendix is to perform the
Passarino-Veltman reduction \cite{pv} of
\eqref{partintegral}, which will lead us to
re-express $\cC (m_1 , m_2)$
in terms of cut-boxes, cut-triangles and cut-bubbles.

The explicit forms for the Dirac traces
involve Lorentz contractions over the various momenta,
so in a short-hand notation we can write these as
\beq
\label{lazy}
T(i,j,m_1)_{\mu}\, l_1^{\mu}
\ := \
{\tr}_{+}(\kslash_i \kslash_j
\kslash_{m_1} \lslash_1)
\ .
\eeq
$\cC (m_1 , m_2) $ can then be recast as
\beq
\label{partintegralii}
\cC (m_1 ,m_2)  \ = \
{T(i,j,m_1)_{\mu} \, T(i,j,m_2)_{\nu} \over (i\cdot j)^2}
%\,
%\int\!\frac{dz_1}{z_1}\frac{dz_2}{z_2}
\ \cI^{\m \n} (m_1, m_2, P_{L;z})
%d{\rm LIPS}(l_2,-l_1;P_{L;z})\frac{
%l_1^{\mu}\,
%l_2^{\nu}}
%{(m_1\cdot l_1)(m_2\cdot l_2)}
\ ,
\eeq
where%
\footnote{For the rest of this Appendix we drop the subscript
$z$ in $P_{L;z}$ for the sake of brevity.}
\beq
\label{imunu}
\cI^{\mu\nu} (m_1, m_2, P_{L})
\ = \
\int\! d{\rm LIPS}(l_2,-l_1;P_{L})
\,
\frac{l_1^{\mu}\, l_2^{\nu}}{
%(i\cdotj)^2
(m_1\cdot l_1)(m_2\cdot l_2)}
\ .
\eeq
$\cI^{\mu\nu} (m_1, m_2, P_{L})$
%corresponds to a tensor box diagram, and as such it
contains  three independent momenta $m_1$, $m_2$ and  $P_L$.
On general grounds, we can therefore decompose it as
\begin{eqnarray}
\label{ansatz}
\cI^{\mu\nu} &=& {\eta}^{\mu\nu}\, \cI_0
\ +\
m_1^{\mu}m_1^{\nu}\, \cI_1
\ + \
m_2^{\mu}m_2^{\nu}
\, \cI_2
\ + \
P_L^{\mu}P_L^{\nu}\, \cI_3\ + \
m_1^{\mu}m_2^{\nu}\cI_4
\nonumber\\ \cr
&+ &
m_2^{\mu}m_1^{\nu}\, \cI_5
\ + \ m_1^{\mu}P_L^{\nu}\, \cI_6
\ + \
P_L^{\mu}m_1^{\nu}\, \cI_7
\ + \ m_2^{\mu}P_L^{\nu}\, \cI_8
\ + \ P_L^{\mu}m_2^{\nu}\, \cI_{9}
\ ,
\end{eqnarray}
for some coefficients $\cI_i, i=0,...9$. One can
then contract with different combinations of
the independent momenta in order to solve for the
$\cI_{i}$.
For instance, two of the integrals that
we will end up having to do are
${\eta}^{\mu\nu}\cI_{\mu\nu}$
and $m_1^{\mu}m_1^{\nu}\cI_{\mu\nu}$.
Using momentum conservation $l_2-l_1+P_L=0$
and the identity $a\cdot b = {(a+b)^2}/2=-{(a-b)^2}/2$
for $a$,$b$ massless momenta,
we can convert these integrals into
ones which have the general form
\beq
\label{genform}
{\tilde{\cI}}^{(a,b)}=\int \frac{d{\rm
LIPS} (l_2, -l_1; P_L)}{(l_1\cdot m_1)^{a}(l_2\cdot m_2)^{b}}
\ ,
\eeq
possibly with a kinematical-invariant coefficient,
and with $a$ and $b$
ranging over the values $1,0,-1$.
The results of these integrals are collected in
Appendix B.
As an example, we find that
\beq
\label{point}m_1^{\mu}m_1^{\nu}\cI_{\mu\nu}\ = \
\int\! d{\rm LIPS}(l_2 , -l_1 ; P_{L}) \,
\frac{(l_1\cdot m_1)}{(l_2\cdot m_2)}
\,  - \,
(m_1\cdot P_L)
\int\!
{d{\rm LIPS}(l_2 , -l_1 ; P_{L})  \over (l_2\cdot P_L)}
\ .
\eeq
Considering the values $(a,b)$, the case $(1,1)$ 
is a cut scalar box,
$(1,0)$ and $(0,1)$ are cut scalar triangles, 
$(1,-1)$ and $(-1,1)$ are cut
vector triangles, whilst $(0,0)$ is a cut scalar bubble.

Because of the structure of
$T(i,j,m_1)_{\mu}$ and $T(i,j,m_2)_{\nu}$,
terms with coefficients such as
$T(i,j,m_1)_{\mu}T(i,j,m_2)_{\nu}{m_1}^{\mu}{m_2}^{\nu}$
are zero, and thus some of the $\cI_{i}$
do not contribute to the final answer.
The only contributing terms are found to be
$\cI_3$, $\cI_5$,  $\cI_7$ and $\cI_8$, and we find that
\begin{eqnarray}
\label{reducedintegral}
\cC (m_1 , m_2)  &=& \frac{{\tr}_{+}(\kslash_i
\kslash_j \kslash_{m_1} \Pslash_L)
{\tr}_{+}(\kslash_{i}\kslash_{j}\kslash_{m_2}
\Pslash_{L})}{(i\cdot j)^2}
\, \cI_3
%\, + \,
\nonumber \\
&+ &
\frac{{\tr}_{+}(\kslash_i \kslash_j \kslash_{m_1}
\kslash_{m_2}){\tr}_{+}(\kslash_{i}\kslash_{j}
\kslash_{m_2}\kslash_{m_1})}{(i\cdot j)^2}
\, \cI_5
\nonumber \\
&+ & \frac{{\tr}_{+}(\kslash_i \kslash_j
\kslash_{m_1} \Pslash_L){\tr}_{+}(\kslash_{i}
\kslash_{j}\kslash_{m_2}\kslash_{m_1})}{(i\cdot j)^2}
\, \cI_7
\nonumber \\
&+ &
%\, + \,
\frac{{\tr}_{+}(\kslash_i \kslash_j
\kslash_{m_1} \kslash_{m_2}){\tr}_{+}(\kslash_{i}
\kslash_{j}\kslash_{m_2}\Pslash_{L})}{(i\cdot j)^2}\,
\cI_8
\ .
\end{eqnarray}
The inversion of (\ref{ansatz}) in order to find 
the coefficients
is tedious and somewhat lengthy, so we just
present the results for the relevant
$\cI_i$ in (\ref{reducedintegral}) above:
\begin{eqnarray}
\label{longeqn1}
\cI_3 &=& \frac{1}{N^2}
\left\{2(m_1\cdot m_2)P_L^2\,
{\tilde{\cI}}^{(0,0)}-N(m_1\cdot P_L)\,
{\tilde{\cI}}^{(1,0)}+N(m_2\cdot P_L)\,
{\tilde{\cI}}^{(0,1)}\right.{}
\nonumber \\
&+ & \left. 2(m_2\cdot P_L)^2\,
{\tilde{\cI}}^{(-1,1)}+2(m_1\cdot P_L)^2\,
{\tilde{\cI}}^{(1,-1)}\right\}
\ ,
\\ \cr
\label{longeqn2}
\cI_5 &=& \frac{1}{(m_1\cdot m_2)^2N^2}
\Bigg\{\bigg[
4(m_1\cdot P_L)^2(m_2\cdot P_L)^2-6(m_1\cdot P_L)
(m_2\cdot P_L)(m_1\cdot m_2)P_L^2
\nonumber \\
&+ &
3(m_1\cdot m_2)^2\left(P_L^2\right)^2\bigg]\,
{\tilde{\cI}}^{(0,0)}+\left[2(m_1\cdot P_L)^2(m_2\cdot P_L)-
\frac{3}{2}(m_1\cdot m_2)P_L^2\right] N(m_1\cdot P_L)\,
{\tilde{\cI}}^{(1,0)}
\nonumber \\
&-  &
\left[2(m_1\cdot P_L)^2(m_2\cdot P_L)-
\frac{3}{2} (m_1\cdot m_2)P_L^2\right]
N(m_2\cdot P_L)\,  {\tilde{\cI}}^{(0,1)}+\frac{N^3}{4}\,
{\tilde{\cI}}^{(1,1)}
\nonumber \\
&+ &
2\bigg[(m_1\cdot m_2)P_L^2-
(m_1\cdot P_L)(m_2\cdot P_L)\bigg]
(m_2\cdot P_L)^2\,  {\tilde{\cI}}^{(-1,1)}
\nonumber \\ &+ &
2\bigg[(m_1\cdot m_2)P_L^2- (m_1\cdot P_L)(m_2\cdot P_L)\bigg]
(m_1\cdot P_L)^2\, {\tilde{\cI}}^{(1,-1)}\Bigg\}
\ ,
\\ \cr
\label{longeqn3}
\cI_7 &=& \frac{1}{(m_1\cdot P_L)(m_1\cdot m_2)N^2}
\Bigg\{\bigg[2(m_1\cdot P_L)^2(m_2\cdot P_L)^2-
3(m_1\cdot P_L)(m_2\cdot P_L)(m_1\cdot m_2)P_L^2\bigg]
\nonumber \\
&\cdot &
{\tilde{\cI}}^{(0,0)}+
\frac{1}{2}(m_1\cdot m_2)P_L^2N(m_1\cdot P_L)\,
{\tilde{\cI}}^{(1,0)}-(m_1\cdot P_L)
N(m_2\cdot P_L)^2\, {\tilde{\cI}}^{(0,1)}
\nonumber \\
&- &
2(m_1\cdot P_L)(m_2\cdot P_L)^3\,
{\tilde{\cI}}^{(-1,1)}-(m_1\cdot m_2)P_L^2
(m_1\cdot P_L)^2\, {\tilde{\cI}}^{(1,-1)}\Bigg\}
\ ,
\\ \cr
\label{longeqn4}
\cI_8 &=&
\frac{1}{(m_2\cdot P_L)(m_1\cdot m_2)N^2}
\Bigg\{\bigg[2(m_1\cdot P_L)^2(m_2\cdot P_L)^2-
3(m_1\cdot P_L)(m_2\cdot P_L)(m_1\cdot m_2)P_L^2\bigg]
\nonumber \\ & \cdot &
{\tilde{\cI}}^{(0,0)}+
(m_2\cdot P_L)N(m_1\cdot P_L)^2\,
{\tilde{\cI}}^{(1,0)}-
\frac{1}{2}(m_1\cdot m_2)P_L^2N(m_2\cdot P_L)\,
{\tilde{\cI}}^{(0,1)}
\nonumber \\
&- & (m_1\cdot m_2)P_L^2(m_2\cdot P_L)^2\,
{\tilde{\cI}}^{(-1,1)}-2(m_1\cdot P_L)^3(m_2\cdot P_L)\,
{\tilde{\cI}}^{(1,-1)}\Bigg\}
\ ,
\end{eqnarray}
where
$N = (m_1\cdot m_2)P_L^2-2(m_1\cdot P_L)(m_2\cdot P_L)$.
The explicit expressions for the relevant 
${\tilde{\cI}}^{(a,b)}$
are summarised in Appendix B.

Combining \eqref{reducedintegral} and
\eqref{longeqn1}-\eqref{longeqn4}
with the identity \eqref{remarkable}
and the explicit expressions for the integrals
$\tilde{\cI}^{(a,b)}$ in Appendix B,
we arrive at the final result \eqref{partNeq1cut}.

%%%%%%%%%%%%%%%%%%%%%%%%%%%%%%%%%%%%%%%%%%%%%%%%%%%%%%%%%%%
\Appendix{Box and triangle discontinuities
from phase-space integrals}
The integrals that arise in the Passarino-Veltman reduction
in Appendix A  have the general form:
\beq
\label{ab}
{\tilde{\cI}}^{(a,b)} =
\int \!\frac{d^{4-2\epsilon}{\rm LIPS} (l_2 , -l_1 ; P_{L;z})}
{(l_1\cdot m_1)^{a}(l_2\cdot m_2)^{b}}
\ ,
\eeq
where we have introduced dimensional regularisation in
dimension $D=4-2\epsilon$
\cite{'tHooft:fi} in order to deal with infrared 
divergences.

There are six cases to deal with:
${\tilde{\cI}}^{(0,0)}$,
${\tilde{\cI}}^{(1,0)}$,
${\tilde{\cI}}^{(0,1)}$,
${\tilde{\cI}}^{(1,1)}$,
${\tilde{\cI}}^{(-1,1)}$,
${\tilde{\cI}}^{(1,-1)}$,
though due to symmetry we can transform
${\tilde{\cI}}^{(1,0)}$ into
${\tilde{\cI}}^{(0,1)}$, and
${\tilde{\cI}}^{(-1,1)}$ into
${\tilde{\cI}}^{(1,-1)}$,
so we only need consider four cases
overall.

Generically we will evaluate these integrals 
in convenient special frames
following Appendix B of \cite{bst}, 
with a convenient choice for
$m_1$ and $m_2$. For instance, in the case of
${\tilde{I}}^{(1,1)}$ it is convenient 
to transform to the centre
of mass frame of the vector $l_1-l_2$, so that
\beq
\label{comass}
l_1 \ = \ \frac{1}{2}P_{L;z} \bigl(  1 \, , \, {\bf v} \big)\ ,
\qquad l_2 \ = \   \frac{1}{2}P_{L;z} \bigl( -1 \, , \,  {\bf
v}\big), \eeq and write \beq \label{comass2}
{\bf v} \ = \
(\sin\theta_1\cos\theta_2\, , \, \ldots \, , \,   \cos\theta_1)
\ .
\eeq
Using a further spatial rotation we write
\begin{equation}
\label{comass3}
   m_1 = (m_1,0,0,m_1) \ , \qquad
   m_2 = (A,B,0,C)
\ ,
\end{equation}
with the mass-shell condition $A^2 = B^2 + C^2$.

After integrating over all angular coordinates except $\theta_1$
and $\theta_2$, the two-body phase space measure in 
$4-2\epsilon$ dimensions becomes \cite{bst}
\beq
d^{4-2\epsilon}{\rm LIPS}
(l_2, -l_1; P_{L;z}) \ = \
{ \pi^{ {1\over 2} - \epsilon} \over 4
\, \Gamma \big({1\over 2} - \epsilon\big) } \, \left| P_{L;z}
\over 2 \right|^{- 2 \epsilon} \ d\theta_1\, d\theta_2\, \,
(\sin\theta_1)^{1-2\epsilon} \, (\sin\theta_2)^{-2\epsilon}
\ .
\eeq
As a result of this and of our parametrizations of
$l_1,l_2,m_1$ and $m_2$, the integrals take the form
%reminiscent of those in \cite{vanNeerven:1985xr,wim}, with
\beq
\label{wibble}
\tilde{\cI}^{(a,b)}\ = \
\Lambda^{(a,b)}{ \pi^{ {1\over 2} - \epsilon} \over 4 \,
\Gamma \big({1\over 2} - \epsilon\big) } \, \left| P_{L;z}
\over 2
\right|^{- 2 \epsilon}{\cal{J}}^{(a,b)}\ ,
\eeq
where
\begin{eqnarray}
\label{lambda0}
{\Lambda}^{(0,0)}&=&1
\ ,
\\ \nonumber
\label{lambda1}
{\Lambda}^{(1,0)} &=&
\frac{2}{P_{L;z}m_1}
\ ,
\\ \nonumber
\label{lambda2}
{\Lambda}^{(0,1)}
&=&-\frac{2}{P_{L;z}m_2}
\ ,
\\ \nonumber
\label{lambda3}
{\Lambda}^{(1,1)}&=&
-\frac{4}{P_{L;z}^2m_1}
\ ,
\\ \nonumber
\label{lambda4}
{\Lambda}^{(-1,1)}&=&-m_1
\ ,
\\ \nonumber
\label{lambda5}
{\Lambda}^{(1,-1)}&=&-m_2
\ ,
\end{eqnarray}
%is a kinematical factor depending on
%our particular parametrisation
and
${\cal{J}}^{(a,b)}$ is the angular integral
\beq
\label{bigint}
{\cal{J}}^{(a,b)}\ := \
\int_{0}^{ \pi}\! d\theta_1 \int_{0}^{2
\pi}d\theta_2\,\! \frac{(\sin\theta_1)^{1-2\epsilon}
(\sin\theta_2)^{-2\epsilon} }
 { (1 - \cos\theta_1)^a(A + C\cos\theta_1 +
B \sin\theta_1 \cos\theta_2)^b }
\ .
\eeq
The integrals (\ref{bigint}) have been evaluated in
\cite{vanNeerven:1985xr} for the values of $a$ and $b$ specified
above, and we borrow the results in a form from \cite{wim}:
\begin{eqnarray}
\label{jay1}
{\cal{J}}^{(0,0)}&=&
\frac{2\pi}{1-2\epsilon}
\ ,
\\ \nonumber
\label{jay2}
{\cal{J}}^{(1,0)}
&=&-\frac{\pi}{\epsilon}
\ ,
\\ \nonumber
\label{jay3}{\cal{J}}^{(1,1)}
&=&
-\frac{\pi}{\epsilon}\frac{1}{A}
\ {}_2F_1\left(1,1,1-\e,\frac{A-C}{2A}\right)
\ ,
\\ \nonumber
\label{jay4}
{\cal{J}}^{(-1,1)}
&=&
-\frac{2\pi(1-\epsilon)}
{\epsilon(1-2\epsilon)} \ {}_2F_1\left(-1,1,1-\e,
\frac{A-C}{2A}\right)
\ .
\end{eqnarray}
Here, $A$ and $B$ will differ depending on which case we are
considering and our particular parametrization for it, 
but in all
cases the combinations that arise 
can be re-expressed in terms of
Lorentz-invariant quantities using 
suitable identities. In the
case of ${\cal{J}}^{(1,1)}$ for example, 
one uses the easily verified
identities
\begin{equation}
\label{happy}
N(P_{L;z}) \ = \ -P_{L;z}^2(A+C)m_1\, ,
\quad m_1\cdot m_2 = m_1(A-C)
\ , 
\
\end{equation}
where $N(P_{L;z})$ was defined in \eqref{N}.

Eventually, after re-expressing $A$ and $B$ in this way,
and upon application of some standard hypergeometric identities
we find the following:
\begin{eqnarray}
\label{beebop1}
\l^{-1}\, {\tilde{\cI}}^{(0,0)}&=&\frac{2\pi}{1-2\epsilon}
\ ,
\\ \nonumber
\label{beebop2}
\l^{-1}\, {\tilde{\cI}}^{(1,0)}&=&
- \frac{1}{\epsilon}\, \frac{2\pi}{m_1\cdot
P_{L;z}}
\ ,
\\ \nonumber
\label{beebop3}
\l^{-1}\, {\tilde{\cI}}^{(0,1)}
&=&
\frac{1}{\epsilon}\, \frac{2\pi}{m_2\cdot
P_{L;z}}
\ ,
\\ \nonumber
\label{beebop4}
\l^{-1}\, {\tilde{\cI}}^{(1,1)}
&=&
-\frac{8\pi}{N(P_{L;z})}\left\{\frac{1}{\epsilon}+
\log{\left(1-\frac{(m_1\cdot
m_2)
P_{L;z}^2}{N(P_{L;z})}\right)}+
\mathcal{O}({\epsilon})\right\}
\ ,
\\ \nonumber
\label{beebop5}
\l^{-1}\, {\tilde{\cI}}^{(-1,1)}&=&\frac{\pi}{(m_1\cdot
P_{L;z})^2}\left\{-\frac{N(P_{L;z})}{\epsilon}\, \right.
\nonumber
\\ \nonumber
&+ &
\left. \frac{2}{1-2\epsilon}\bigl[(m_1\cdot P_{L;z})(m_2\cdot
P_{L;z})-(m_1\cdot m_2)P_{L;z}^2\bigr]\right\}
\ ,
\\ \nonumber
\label{beebop6}
\l^{-1}\, {\tilde{\cI}}^{(1,-1)}&=&\frac{\pi}{(m_2\cdot
P_{L;z})^2}\left\{-\frac{N(P_{L;z})}{\epsilon}
\right.
\nonumber \\
&+ &
\left.
\frac{2}{1-2\epsilon}
\bigl[ (m_1\cdot P_{L;z})(m_2\cdot
P_{L;z})-(m_1\cdot m_2)P_{L;z}^2\bigr]\right\},
\nonumber
\end{eqnarray}
where $\l$ is the ubiquitous factor
\beq
\label{ubi}
\l \ := \
{ \pi^{ {1\over 2} -
\epsilon} \over 4 \, \Gamma \big({1\over 2} - \epsilon\big) } \,
\left| P_{L;z} \over 2 \right|^{- 2 \epsilon}
\ .
\eeq

%%%%%%%%%%%%%%%%%%%%%%%%%%%%%%%%%%%%%%%%%%%%%%%%%%%%%%%%%%%
\Appendix{Reconstructing triangles from dispersion integrals in
a gauge-invariant way}
In this Appendix we find a new representation of
the triangle function
\beq
\label{triangle-funct}
T(p, P, Q) \ = \
{\log (Q^{2} / P^{2}) \over Q^{2 }- P^{2}}
\ ,
\eeq
as the dispersion integral of a sum of two cut-triangles.%
\footnote{For a review of dispersion relations, see
\cite{bib}.} 
A comment on gauge-(in)dependence is in order here.
Recall from Section 2, Eq.~\eqref{off}, that in
the approach of \cite{bst} to loop diagrams
one introduces an arbitrary null vector $\eta$
in order to perform loop integrations.
The corresponding gauge dependence should
disappear in the expression for scattering amplitudes.
In what follows we will work in  an arbitrary gauge,
and show analytically that gauge-dependent terms
disappear in the final result for the triangle function.
Perhaps unsurprisingly, this gauge invariance
will also hold for the finite-$\epsilon$ version
of $T(p, P, Q)$, which we define in
\eqref{epstriangle}.

To begin with, recall from \eqref{listofct}
that the basic quantity we have to compute reads
\beq
\label{crcr}
\cR\ := \
\int\! {dz \over z} \left[
{(P_{z}^{2})^{-\epsilon}\over ( P_{z}p) } \, + \,
{(Q_{z}^{2})^{-\epsilon}\over (Q_{z}p)}
\right]
\ ,
\eeq
where $P+Q+p=0$.
We will work in an arbitrary gauge, where
\beq
P_{z}\ := \ P \, - \, z \eta \ ,
\qquad
Q_{z} \  := \ Q \, + z \eta
\ .
\eeq
A short calculation shows that
\beqa
P_{z}p & =  & Pp \Big[
1 - b_{P } (P^{2 } - P_{z}^{2})
\Big]
\ ,
\\
Q_{z}p & = & Qp \Big[
1 - b_{Q} (Q^{2} - Q_{z}^{2})
\Big]
 \ ,
 \eeqa
where
\beq
b_{P} \ := \
{\eta p \over 2 (\eta P) (pP)}
\ ,
\qquad
b_{Q} \ := \
{\eta p \over 2 (\eta Q)(pQ)}
 \ .
 \eeq
It is also useful to notice the relation
\beq
\label{bpbq}
{1\over b_{Q}}
\ = \
{1\over b_{P}} \, + \, Q^{2} -P^{2}
\ ,
\eeq
as well as $(Pp)=-(Qp)= (1/2)(Q^{2} - P^{2})$,
which trivially follows from momentum conservation.
We can then rewrite \eqref{crcr} as
\beq
\cR \ = \ \cI_{1  } \, - \, \cI_{2}
\ ,
\eeq
where
\beqa
\cI_{1} & :=  &
{1\over (Pp)}
\int\! ds' \, (s')^{-\epsilon }\
{1\over (s'-P^{2}) \big[
1-b_{P} (P^{2 } - s')
\big]}
\\ \nonumber
&=&
{\pi \csc  (\pi \epsilon) \over (Pp)}
\left[
(-P^{2})^{-\epsilon} \, - \,
\left(
{-b_{P}\over b_{P}P^{2} -1}
\right)^{\epsilon}
\right]
\ ,
\eeqa
\beqa
\cI_{2} &:= &
{1\over (Pp)}
\int\! ds' \, (s')^{-\epsilon }\
{1\over (s'-Q^{2}) \big[
1-b_{Q} (Q^{2 } - s')
\big]}
\\ \nonumber
&=&
{\pi \csc  (\pi \epsilon) \over (Pp)}
\left[
(-Q^{2})^{-\epsilon} \, - \,
\left(
{-b_{Q}\over b_{Q}Q^{2} -1}
\right)^{\epsilon}
\right]
\ .
\eeqa
But \eqref{bpbq} implies
\beq
{-b_{P}\over b_{P}P^{2} -1} \ = \
{-b_{Q}\over b_{Q}Q^{2} -1}
\ ,
\eeq
so that we can finally recast \eqref{crcr} as:
\beq
\label{finaltriang}
\cR \ = \
2\bigl[\pi \epsilon \csc (\pi \epsilon)\bigr]
\,
{1\over \epsilon}
{(-P^{2})^{{-\epsilon}} \, - \,
(-Q^{2})^{-\epsilon}
 \over Q^{2} - P^{2}}
\ = \
2 \bigl[\pi \epsilon \csc (\pi \epsilon) \bigr]
\, T_{\epsilon}(p, P, Q)
\ ,
\eeq
where the $\epsilon$-dependent triangle function
is%
\footnote{The $\epsilon$-dependent triangle function
already appeared in \eqref{epstriangle}.}
\beq
\label{epstriangleagain}
T_{\epsilon} (p, P, Q) \ := \
{1\over \epsilon}
{ (-P^2)^{-\epsilon} - (-Q^2)^{-\epsilon} \over Q^2 - P^2}
\ .
\eeq
This is the result we were after.
Notice that  all the gauge dependence,
i.e.~any dependence on the arbitrary
null vector $\eta$, has completely cancelled out
in \eqref{finaltriang}.

We now discuss the $\epsilon \to 0$ limit of the
final expression \eqref{finaltriang}.
As already discussed in Section 2
(see \eqref{T1} and \eqref{T2}), in studying the
$\epsilon \to 0$ limit of $\cR$  (and hence of
$ T_{\epsilon}(p, P, Q)$)
we need to distinguish the case where $P^2$ and $Q^2$
are both nonvanishing from the case where one of the two,
say $Q^2$, vanishes.
In the former case, we get precisely the triangle function
$T(p,P,Q)$ defined in \eqref{triangle-funct}:
\beq
\lim_{\epsilon \to  0} \cR \ = \
 2 \, T(p, P, Q)\, , \qquad
P^2 \neq 0 \, , \, Q^2 \neq 0
\ .
\eeq
In the latter case, where $Q^2 =0$, we have instead
\beq
\lim_{\epsilon \to  0} \cR \ = \
-{2\over \epsilon} \,
{(-P^2)^{-\epsilon}  \over P^2}
\, , \qquad
P^2 \neq 0 \, , \, Q^2 = 0
\ ,
\eeq
which corresponds to a degenerate triangle.

The final issue is that of the gauge invariance of the
contributions to the amplitude from the box functions $B$ 
(this is also relevant to the issue 
of gauge invariance
in the $\cN=4$ calculation of \cite{bst}, 
and in that paper we also gave a
general argument for gauge invariance).
We expect that an explicit analytic proof of the
gauge invariance of the box function contribution 
to the amplitude
 could be constructed using identities such as those in
Appendix B of \cite{bst}.
In the meantime, numerical tests have shown that 
gauge invariance is present \cite{Lance}.
Indeed, it would be surprising if this 
were not the case, given that
the correct, gauge invariant, amplitudes 
are derived with the choices
of gauge we have made here and in \cite{bst}.
We have also carried out the MHV diagram analysis of this paper using 
the alternative gauge choice $\eta = k_{m_2}$; one obtains \eqref{Neq1}.

%%%%%%%%%%%%%%%%%%%%%%%%%%%%%%%%%%%%%%%%%%%%%%%%%%%%%%%%%%
\Appendix{Spinor and Dirac-trace identities} 
In this appendix we
present some identities that are needed for many of the
manipulations required to massage the $\mathcal{N}=1$ amplitude
into the form given by BDDK (\ref{BDDKNeq1}).

For spinor manipulations, the Schouten identity is very useful:
\beq
\label{schouten}
\langle i\, j\rangle\langle k\, l\rangle =
\langle i\, k\rangle\langle j\, l\rangle + \langle i\,
l\rangle\langle k\, j\rangle \ .
\eeq
Furthermore we have:
\begin{eqnarray}
\label{spinor1}
\langle i\, j\rangle\left[j\,
i\right]
&=&
{\tr}_{+}(\kslash_i \kslash_j)=2(k_i\cdot k_j)
\ ,
\\ \cr
\label{spinor2}
\langle i\, j\rangle\left[j\, l\right]\langle l\,
m\rangle\left[m\, i\right]
&=&
{\tr}_{+}(\kslash_i \kslash_j
\kslash_{l} \kslash_m)
\ ,
\\ \cr
\label{spinor3}
\langle i\, j\rangle\left[j\, l\right]\langle l\,
m\rangle\left[m\, n\right]\langle n\, p\rangle\left[p\, i\right]
&=&
{\tr}_{+}(\kslash_i \kslash_j \kslash_{l} \kslash_m \kslash_n
\kslash_p)\ ,
\end{eqnarray}
for momenta
$k_i,k_j,k_l,k_m,k_n,k_p$.
For a nice introduction to 
the spinor helicity formalism, see \cite{dixon}.

For dealing with Dirac traces, the following identities are
useful:
\begin{eqnarray}
\label{trace1} {\tr}_{+}(\kslash_i \kslash_j \kslash_{l}
\kslash_m)
&=&
{\tr}_{+}(\kslash_m \kslash_l \kslash_{j}
\kslash_i)
\ ,
\\ \cr
\label{trace2}
{\tr}_{+}(\kslash_i \kslash_j \kslash_{l}
\kslash_m)
&=&
4(k_i\cdot k_j)(k_l\cdot k_m)-{\tr}_{+}(\kslash_j
\kslash_i \kslash_{l} \kslash_m)
\ ,
\end{eqnarray}
for similarly generic momenta.
If $k_i,k_j,k_{m_1}$ and $k_{m_2}$
are massless, while $P_L$ is not necessarily so,
then we have the remarkable identity:
\begin{eqnarray}
\label{remarkable}
2(k_{m_1}\cdot k_{m_2}){\tr}_{+}(\kslash_i
\kslash_j \kslash_{m_1} \Pslash_L){\tr}_{+}(\kslash_i \kslash_j
\kslash_{m_2} \Pslash_L)
%&+&
\nonumber
\\  [3pt] +
P_L^2\,{\tr}_{+}(\kslash_i \kslash_j \kslash_{m_1}
\kslash_{m_2}){\tr}_{+}(\kslash_i
\kslash_j \kslash_{m_2} \kslash_{m_1})
%&-&
\nonumber\\ [3pt]
-2(k_{m_1}\cdot P_L){\tr}_{+}(\kslash_i \kslash_j \kslash_{m_1}
\kslash_{m_2}){\tr}_{+}(\kslash_i \kslash_j \kslash_{m_2}
\Pslash_L)
%&-&
\nonumber \\ [3pt]
-2(k_{m_2}\cdot P_L){\tr}_{+}(\kslash_i
\kslash_j \kslash_{m_1} \Pslash_L){\tr}_{+}(\kslash_i \kslash_j
\kslash_{m_2} \kslash_{m_1})
&=&0
\ .
\end{eqnarray}

%%%%%%%%%%%%%%%%%%%%%%%%%%%%%%%%%%%%%%%%%%%%%%%%%%%%%%%%%%%

\end{document}